\documentclass[aps,prd,amsmath,amssymb,showpacs]{revtex4-2}

\usepackage{epsfig,float}
\usepackage{graphicx}
\usepackage{dcolumn}
\usepackage{morefloats}
\usepackage{color}
\usepackage{slashed}
\usepackage{bbm}
\usepackage{bm}
\usepackage{amsmath}
\usepackage{multirow}
\usepackage[colorlinks=true,linkcolor=blue]{hyperref}
\usepackage{multirow}
\usepackage[T1]{fontenc}
\usepackage{dsfont}
\usepackage{comment}
   
\usepackage{subfigure}
\usepackage{appendix}

\begin{document}

\title{ Study of the decay pattern of $f_0 (1370)$ as a $\kappa \bar{\kappa }$ molecular state }

\author{Yin Cheng $^{1}$~\footnote{Email: chengyin@itp.ac.cn} and Bing-Song Zou$^{2}$~\footnote{Email: zoubs@mail.tsinghua.edu.cn} }

\affiliation{ 1) Institute of Theoretical Physics,
        Chinese Academy of Sciences, Beijing 100190, China}

\affiliation{ 2) Department of Physics and Center for High Energy Physics, Tsinghua University, Beijing 100084, China}

\begin{abstract}

Under the hypothesis that the $f_0(1370)$ is a $\kappa \bar{\kappa}$ molecular state, we calculate the partial widths of its various decay channels, 
including the two-body decay $K \bar{K}$, $\pi \pi$, $\eta \eta$ and the four-body decay $\rho \rho / \sigma \sigma \to 4 \pi$ and $K \bar{K} \pi \pi$. 
Treating the coupling $g_{f_0(1370) \kappa \bar{\kappa}}$ as a free parameter, 
we find that a value of $25 \sim 40$ GeV brings the total width of $f_0(1370)$ in line with the measured value $200\sim 500$ MeV.
 At the center-of-mass energy $\sqrt{s}=1.37$ GeV, 
 the channels that mainly contribute to the total width are $K \bar{K}$, 
 $\pi \pi$ and $4 \pi$ ranked as $\Gamma(K \bar{K }) > \Gamma(4 \pi) > \Gamma (\pi \pi) $ with $g_{f_0(1370) \kappa \bar{\kappa}}= 35$ GeV. Around $1.37$ GeV, the decay widths of the two-body channels $K \bar{K}$, $\pi \pi$ and $\eta \eta$ remain stable with variation in $\sqrt{s}$, whereas the decay widths of the four-body channels $4 \pi$ and $K \bar{K }\pi \pi$ increase continuously with $\sqrt{s}$.  
Most current data are model dependent and conflicting, particularly regarding the conclusion of $4 \pi$ dominance and the ratio of $K\bar{K}$ to $\pi \pi$ decay widths. The current data can not rule out the $\kappa \bar{\kappa}$ assignment for $f_0(1370)$. Further reliable theoretical and experimental analyses of $f_0(1370)$ are required to reveal its nature. 

\end{abstract}

\maketitle
\section{Introduction}

Quantum Chromodynamics (QCD) was established as the theory of strong interaction almost 50 years ago. However, its low-energy regime is still not well understood due to the breakdown of perturbation theory, particularly the scalar spectrum below $2$ GeV. The strange mass pattern of the lightest scalars $f_0(500)$, also called $\sigma$, $a_0(980)$, $f_0(980)$ contradicts the quark model description, which raises some exotic assignment for them, such as molecular state and tetraquark state~\cite{Baru:2003qq,tHooft:2008rus,Oller:1997ti,Jaffe:1976ig,Pelaez:2003dy,Barnes:1985cy,Close:1992ay}.
It seems that the scalars above $1$ GeV should be assigned to the lowest $0^{++}$ $q\bar{q}$ nonet,
which are expected to be heavier than the lowest pseudoscalars and vectors due to its $^3P_0$ configuration.
 However, the set of particles possibly filled into this nonet seems to be supernumerary and their mass ordering is also anomalous as shown in the Review of Particle Physics (RPP)~\cite{ParticleDataGroup:2024cfk}.
  The properties of the $f_0(1500)$-$f_0(1370)$ are incompatible with them being the singlet-octet system in $q\bar{q}$ nonet. 
  If $f_0(1370)$ is considered as dominantly $n\bar{n}$ state then the $s\bar{s}$ state is expected to be $200 \sim 300$ MeV heavier and decay strongly into $K\bar{K}$. 
  While the partial widths of $f_0(1500)$ decaying into $\pi \pi$ and $4 \pi$ are dominant.
   Its decay ratio to $K\bar{K}$ is only $8.5 \pm 1.0 \% $ as reported in the RPP~\cite{ParticleDataGroup:2024cfk}.

In light of this phenomenology, earlier Amsler and Close~\cite{Amsler:1995td,Amsler:1995tu} suggested that the scalars $f_0(1370)$,
 $f_0(1500)$ and $f_0(1710)$ can originate from a mixing scheme of the scalar glueball with the scalar $q\bar{q}$ state. 
 Subsequently, numerous works have been devoted to the different glueball mixing schemes~\cite{Close:2005vf,Giacosa:2005zt,Chatzis:2011qz,Vento:2015yja}.


On the other hand, some studies on the meson-meson interactions suggest that the $f_0(1370)$ may be dynamically generated in the isoscalar $\rho \rho$ $S$-wave interaction~\cite{Molina:2008jw, Gulmez:2016scm, Geng:2008gx}.
However, when the $\pi \pi$ and $K\bar{K}$ coupled-channels are included, Ref.~\cite{Wang:2019niy} claims that the $\rho \rho$ resonance aligns more closely with the $f_0(1500)$, based on its pole position and partial decay widths.
Assigning the $f_0(1500)$ and $f_0(1710)$ as $\rho \rho $ and $K^* \bar{K}^*$ molecular states,
respectively, Ref.~\cite{Yang:2024wzp} has investigated their plausible isospin partners $a_0(1450)$ and $a_0(1710)$.
In the present work, we use the term ``molecule'' loosely to refer to a state whose formation is governed by a dominant meson-meson component,
 rather than requiring the strict weak-binding condition of a bound state lying just below the corresponding threshold.

It is worth mentioning that the signal of $f_0(1370)$ in experiments is subtle and controversial.
 In 1966, a measurement of the $4\pi$ spectrum in the $\bar{p}n \to 2\pi^+ 3\pi^-$ process~\cite{Bettini:1966zz} found a deviation from the phase space distribution, 
 which is explained by a scalar $\rho \rho$ resonance with mass $m=1.41$ GeV and width $\Gamma=90$ MeV.  
 RPP lists this data as an observation of $f_0(1370)$~\cite{ParticleDataGroup:2024cfk}.
 The width in this analysis is too narrow compared to those in other analyses of $f_0(1370)$, which is about $300$ MeV,
but it is consistent with the average width of $f_0(1500)$. 
 A reanalysis of $\bar{p}n \to 2 \pi^+ 3\pi^-$ by Gaspero~\cite{Gaspero:1992gu} shows that this reaction is dominated by a $4 \pi$ resonance with mass $m=1386$ MeV, width $\Gamma=310$ MeV and quantum number $I^G J^{PC}=0^+ 0^{++}$,
which decays to both $\rho^0 \rho^0$$(29\%)$ and $\sigma \sigma$$(71)\%$.
A compatible scalar was observed in $p\bar{p} \to \pi^+ \pi^- 3 \pi^0$ by Crystal Barrel Collaboration~\cite{CrystalBarrel:1994doj}, which decay through both $\rho^+ \rho^-$ and $\sigma\sigma$ intermediate states with different relative strengths for these two decay modes. 
 The discrepancy is natural because the comparison is made between the ratios of $\rho^0 \rho^0$/$\sigma \sigma$ and $\rho^+ \rho^-$/$\sigma \sigma$.
Later measurements of $\bar{p} d \to \pi^- 4 \pi^0 p$ require two scalar states, $f_0(1370)$ and $f_0(1500)$~\cite{CrystalBarrel:2001xud}, with mass and width parameters consistent  with prior results. While analyses of Crystal Barrel data of $4 \pi$~\cite{CrystalBarrel:2001xud,CRYSTALBARREL:2001ldq} indicate strong coupling of the $f_0(1370)$ to the $4 \pi$ final state, studies of its two-body decays have produced highly contradictory results. This is further complicated by the $f_0(1500)$, a $\rho \rho$ molecule candidate that is expected to dominate the $4 \pi$ spectrum. Therefore,
the $f_0(1370)$ signal in the $4 \pi$ spectrum is still ambiguous.

 The Crystal Barrel data in Refs.~\cite{CrystalBarrel:1992tzh,CrystalBall:1994rrl,Amsler:1995gf} manifest a peak in the $\pi \pi $ and $\eta \eta$ $S$-wave at about $1400$ MeV, 
 as well as a further peak at about $1560$ MeV.
Its width varies between $200$ and $700$ MeV, depending on theoretical models.
A combined fit~\cite{Abele:1996si} to the Crystal Barrel data on $p\bar{p}\to 3 \pi^0$, $\eta \eta \pi^0$ and $\eta \pi^0 \pi^0$,
 which is fully consistent with several other set of data: CERN-Munich data on $\pi^- \pi^+ \to \pi^- \pi^+$, GAMS data on $\pi^- p \to \pi^0 \pi^0 n$,
 and data from ANL and BNL on $\pi^- \pi^+ \to K\bar{K}$, claimed that the $f_0(1370)$ is definitely required by the data on $3 \pi^0$ and $\eta \pi^0 \pi^0$,
 as well as $\pi \pi \to K \bar{K}$.

  Klempt and Zaitsev~\cite{Klempt:2007cp, Ochs:2013gi} question the existence of $f_0(1370)$. 
They argue that in the $\eta \eta$ spectra in $p \bar{p}$ annihilation at higher primary energy~\cite{E760:1993bjv,CrystalBarrel:2002qpf,Uman:2006xb}, 
$f_0(1500)$ is always observed clearly but no such ``direct'' evidence of the $f_0(1370)$. 
However, $f_0(1370)$ might have weaker decay rate to the $\eta \eta$ channel compared to $f_0(1500)$, and as stated in~\cite{Uman:2006xb} the resonances near the boundaries of the Dalitz plot suffer from the effect of the anticuts.
Five primary sets of data requiring $f_0(1370)$ are refitted with suitable Breit-Wigner amplitudes by Bugg to further confirm its existence~\cite{Bugg:2007ja}.

The partial waves analyses from $\pi^- p \to K \bar{K} n$ and $\pi^+ n \to K^- K^+ p $, i.e., the $\pi \pi \to K \bar{K}$ scattering, at Argonne National Laboratory (ANL)~\cite{Cohen:1980cq,Polychronakos:1978ur,Martin:1979gm} 
and Brookhaven National Laboratory (BNL)~\cite{Etkin:1981sg} favored the solution that requiring an $f_0$ enhancement around $1300$ MeV.
Moreover, a model-independent dispersive analyses on $\pi \pi $ scattering data claim the confirmation of the existence of $f_0(1370)$~\cite{Pelaez:2022qby}.

For the $p p$ central production, an early measurement of $\pi^+ \pi^-$ system by WA102 Collaboration~\cite{WA102:1999avd}
claimed that it was not possible to describe the data above $1$ GeV without the addition of both the $f_0(1370)$ and $f_0(1500)$ resonances.
Similar results were obtained from a coupled-channel analysis~\cite{WA102:1999fqy} of centrally produced $K^+ K^-$~\cite{WA102:1999nvm} and $\pi^+ \pi^-$~\cite{WA102:1999avd} final states.
 A relatively recent data from STAR Collaboration~\cite{STAR:2020dzd} on $p p$ central production of $\pi^+ \pi^-$ and $K^+ K^-$ 
 also claim the possible contribution from $f_0(1370)$ in addition to the $f_0(1500)$.  
Based on the WA102 measurements in $p p$ central production, the ratio of $f_0(1370) \to K\bar{K}$ to $f_0(1370) \to \pi \pi $ is determined to be $0.46 \pm 0.19$~\cite{WA102:2000lao,WA102:1999fqy}.

For the production in $J/\psi$ radiative decays,
Ref.~\cite{Dobbs:2015dwa} found convincing evidence of the existence of $f_0(1370)$ in $J/\psi$ and $\psi(2S)$ radiative decay, 
although only in the $K\bar{K}$ decay channel.
 Due to limited statistics, partial-wave analysis was not performed in~\cite{Dobbs:2015dwa}. Consequently,
 the $f_0(1370)$ signal in the $\pi \pi$ channel is obscured by the strong excitation of nearby $f_2(1270)$. 
However, partial-wave analyses of higher-statistics $J/\psi \to \gamma \pi^0 \pi^0$~\cite{BESIII:2015rug} and $J/\psi \to \gamma K_S K_S$~\cite{BESIII:2018ubj} data both reveal a clear $0^{++}$ structure around $1.4$ GeV.
A coupled-channel analysis~\cite{Sarantsev:2021ein} including these two data obtain the branching ratios $Br(J/\psi \to \gamma f_0(1370) \to \gamma \pi \pi ) = (3.8 \pm 1) \times 10^{-4} $ 
and $Br(J/\psi \to \gamma f_0(1370) \to \gamma K \bar{K} ) = (1.3 \pm 0.4) \times 10^{-4} $.
The authors in Ref.~\cite{Sarantsev:2021ein} claim that the interference between neighboring states plays a decisive role 
and the change in the sign of the coupling constant in $\pi\pi$ and $K \bar{K}$
decays for $f_0(1500)$ with respect to the $f_0(1370)$ ``background'' has first been noticed in Ref.~\cite{Minkowski:2004xf}.

For the production in $J/\psi$ hadronic decays, BESIII has measured the $\omega K \bar{K}$~\cite{BES:2004zql}, $\omega \pi^+ \pi^-$~\cite{BES:2004mws}, $\phi K \bar{K}$ and $\phi \pi \pi$~\cite{BES:2004twe} spectra. 
The leading-order mechanism of $J/\psi$ hadronic decay to $V(PP)$ suggests that the quark flavors within the recoiled meson $V$ must appear in the resonance of the two pseudoscalars. 
No signals of $f_0(1370)$ or $f_0(1500)$ are observed in the $\omega K \bar{K}$ and $\omega \pi^+ \pi^-$ spectra.
However, same as decaying to $\pi \pi$, a clear signal of $f_0(1370)$ is observed in the case of recoiling against $\phi$. 
A chiral unitary coupled channel study of the meson-meson $S$ wave interaction in
 Ref.~\cite{Albaladejo:2008qa} provides a successful
 description of the $f_0(1370)$ together with the other scalar-isoscalar resonances below
 2~GeV and assigns $f_0(1370)$ as a pure SU(3) octet state $(u\bar{u}+d\bar{d}-2 s\bar{s})/\sqrt{6}$, not mixed with the glueball.
 This assignment would imply an $f_0(1370)$ signal in $J/\psi$ hadronic decays recoiling
 against $\omega$, which is not observed experimentally. 
 Note that the study in Ref.~\cite{Albaladejo:2008qa} does not include the data from $J/\psi$ hadronic decay,
 which could provide a further test of their octet assignment for $f_0(1370)$.

In the $\phi K \bar{K}$ data, there is a shoulder on its upper side that may be fitted by the interference between $f_0(1500)$ and $f_0(1710)$, 
but no conspicuous enhancement of $f_0(1370)$ is observed. Therefore, a ratio of $Br(J/\psi\to \phi f_0(1370)\to \phi K \bar{K}) / Br(J/\psi\to \phi f_0(1370)\to \phi \pi \pi) = 0.08 \pm 0.08$
is presented, which is much lower than those reported by the Review of Particle Physics.
As discussed in Ref.~\cite{BES:2004twe}, this discrepancy arises from the conspicuous signal in the $\pi \pi$ spectrum and the absence of any corresponding peak in the $K^+K^-$ spectrum.   
  
While numerous studies have advocated for the necessity of the $f_0(1370)$, most analyses to date rely on fits using sums of Breit-Wigner functions. This approach often produces reaction-dependent results, a problem particularly acute for the $f_0(1370)$ given its large width and proximity to the $f_0(1500)$ resonance.

Motivated by the phenomenology that the signals of $f_0(980)$, $f_0(1370)$, and $f_0(1790)$ appear successively in the $\pi \pi$ spectrum in $J/\psi$ hadronic decay recoiling against $\phi$,
where the $s\bar{s}$ pair contributes to the $(PP)$ resonance, we argue that the $f_0(1370)$ might be a $\kappa \bar{\kappa}$ molecular state, inspired by the $f_0(980)$ and $f_0(1790)$ being the plausible $K \bar{K}$ and $K^* \bar{K}^*$ molecules, respectively.
In this work, we aim to investigate the decay pattern for the primary decay channels of $f_0(1370)$, including $\pi \pi$, $K \bar{K}$, $\eta \eta$, $4 \pi$, and $K \bar{K} \pi \pi$, 
within the framework where the $f_0(1370)$ is treated as a $\kappa \bar{\kappa}$ molecular state. 
Regardless of whether the components are broad or narrow, they are nearly free particles inside the molecular state in the weak-binding limit.
 Decay through the almost free components implies that the width of the bound state is expected to be of the same order of magnitude as the width of its components.
 This argument is illustrated by the case of the $T^+_{cc}$ discovered by LHCb~\cite{LHCb:2021auc}, whose width is almost saturated by the decay via its constituents $D^* D \to D D \pi$ and is analyzed to be about $48$ keV.
The width of its constituent $D^*$ is $(83.4 \pm 1.8)$ keV~\cite{ParticleDataGroup:2024cfk}, of the same order as the $T^+_{cc}$ width, exemplifying the expected relation between the width of a hadronic molecule and the widths of its components.
 Hence, aligning with this argument the broad state $f_0(1370)$ is reasonably suspected to be a molecular state composed of $\kappa \bar{\kappa}$.
In Sec.~\ref{Sec:formalism}, we present the decay mechanism and the detailed formalism for the calculation.
In Sec.~\ref{Sec:results}, we will present the numerical results as well as the discussions.
A brief summary will be given in Sec.~\ref{Sec:summary}.

\section{Formalism} \label{Sec:formalism}
   

In this section, we present the detailed formalism for calculating the partial decay widths of the $f_0(1370)$ as a $\kappa \bar{\kappa}$ molecular state.
Using the isospin conventions shown in  Appendix~\ref{Appendix:A}, we have 
\begin{eqnarray}
        &&  |\kappa^+ \kappa^- \rangle \qquad \quad  |\frac{1}{2}, \frac{1}{2} \rangle |\frac{1}{2}, -\frac{1}{2} \rangle = \frac{1}{\sqrt{2}}| 1,0 \rangle + \frac{1}{\sqrt{2}}|0 ,0 \rangle , \label{isospin-1}  \\
        && |\kappa^0 \bar{\kappa}^0 \rangle   \qquad   -|\frac{1}{2}, -\frac{1}{2} \rangle |\frac{1}{2},\frac{1}{2} \rangle= -\bigg(\frac{1}{\sqrt{2}}|1,0 \rangle - \frac{1}{\sqrt{2}}| 0,0 \rangle \bigg).  \label{isospin-2}
\end{eqnarray}
According to these isospin projections, the wave function of $f_0(1370)$ has the form of 
\begin{eqnarray}  
      f_0(1370)= \frac{1}{\sqrt{2}} | \kappa^+ \kappa^- \rangle + \frac{1}{\sqrt{2}} | \kappa^0 \bar{\kappa}^0 \rangle. \label{Eq:wavefunction}
\end{eqnarray}
   
Given that the $\kappa$ is a broad resonance often modeled as a $K\pi$ molecule, considering two such broad states to form a molecular state
is counter-intuitive.
However, this system can be essentially regarded as a $(K\pi)(\bar{K}\pi)$ four-body interacting system.
To avoid the complexity of four-body dynamics, we argue that describing the system as a $\kappa \bar{\kappa}$ molecule is a reasonable and practical approximation.
Within this framework, we evaluate whether the predicted decay pattern of the $f_0(1370)$ remains consistent with available experimental data.
We emphasized that the predictions in this paper are inevitably subject to some model dependence.

Given the complexity of determining the coupling between $f_0(1370)$ and its broad $\kappa$ components, we treat $g_{F\kappa\bar{\kappa}}$ ($F$ denotes $f_0(1370)$ for brevity) as a free parameter in our model.
The value of this coupling is taken as a parameter that is fitted to the measured total width of $f_0(1370)$.
 Specifically, based on the wave function the coupling $g_{F \kappa^+ \kappa^-}=g_{F \kappa^0 \bar{\kappa}^0}=(1/\sqrt{2}) g_{F \kappa \bar{\kappa}}$.
 We set the coupling $g_{F \kappa \bar{\kappa}}$ as a real number in this work, since its phase does not matter for the issues we study here.

\begin{figure}
        \centering
       \subfigure[]{\includegraphics[width=2.0 in]{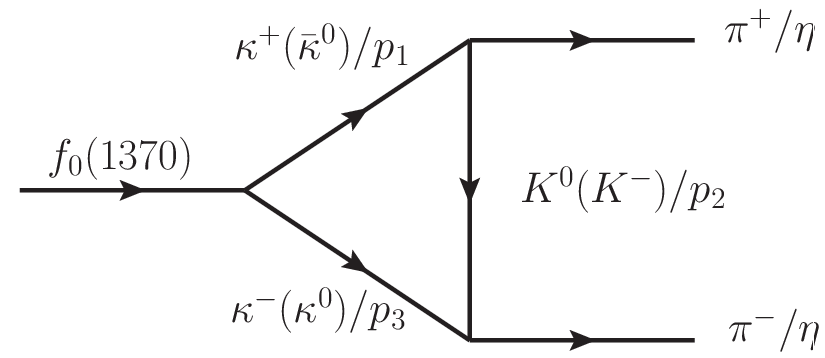}} \qquad \qquad \quad
       \subfigure[]{\includegraphics[width=1.9 in]{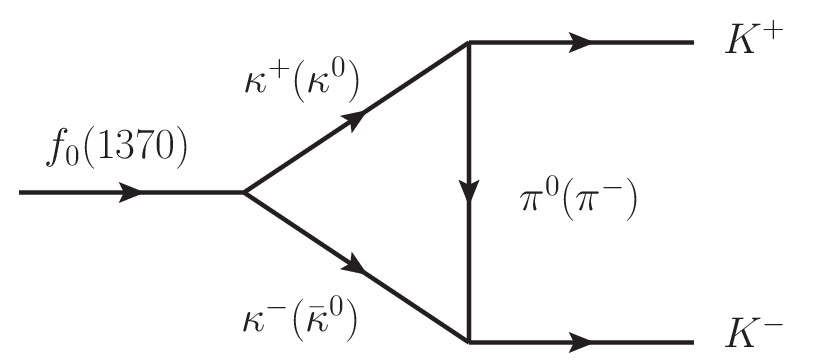}}  \\
       \subfigure[]{\includegraphics[width=2.1 in]{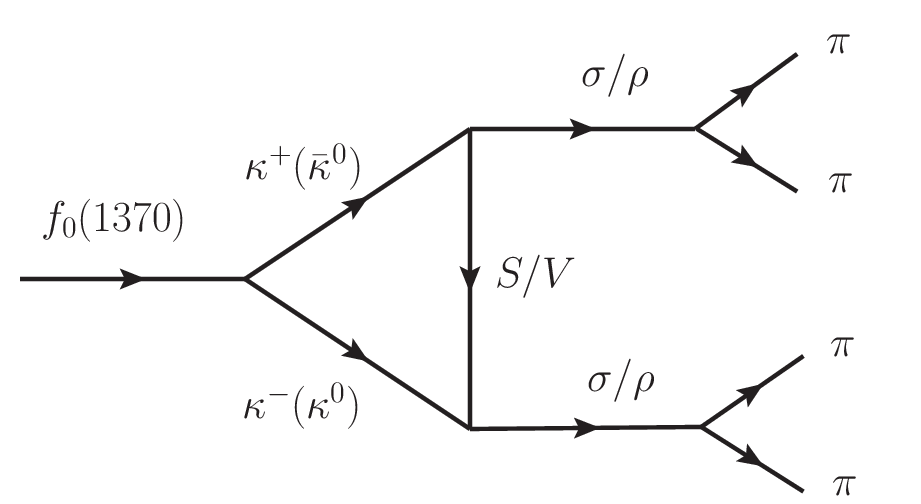}} \qquad \qquad \quad
       \subfigure[]{\includegraphics[width=1.8 in]{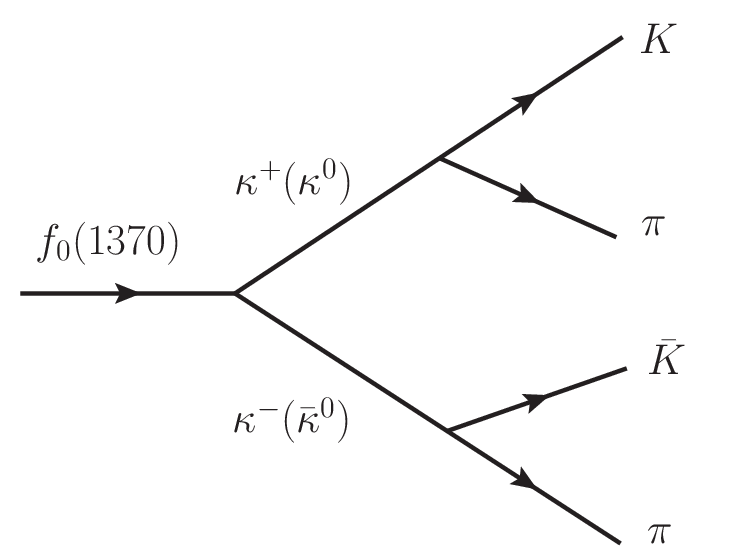}}
        \caption{The mechanism for $f_0(1370)$ decay as a $\kappa \bar{\kappa}$ molecular state.  
        (a) $\pi^+ \pi^-$/$\eta \eta $ final states. (b) $K^+ K^-$ final states. (c) $4 \pi $ final states. $S/V$ denotes that the $t$-channel exchange meson is either a scalar $\kappa$ or a vector $K^*$. (d) $K \bar{K } \pi \pi $ final states.}  \label{Fig:f01370decay}
\end{figure}

In the hadronic molecule picture, the schematic diagrams of $f_0(1370)$ decay to different channels, including $\pi \pi$, $K\bar{K}$, $\eta \eta$, $4\pi$ and $K \bar{K} \pi \pi$, are shown in Fig.~\ref{Fig:f01370decay}. 
Note that the vertices with quantum numbers $SSP$, $SPV$ are forbidden 
and with quantum numbers $SSS$, $SSV$, $SPP$, $SVV$ are allowed based on the parity conservation laws.
For the allowed vertices,
the adopted effective Lagrangians have the forms as
\begin{eqnarray}
        \mathcal{L}_{SSS}&=&g_{SSS} \langle SSS \rangle,  \\
        \mathcal{L}_{SPP}&=&g_{SPP}  \langle SPP \rangle,  \label{Eq:L_SPP} \\
        \mathcal{L}_{SVV}&=& g_{SVV} \langle V^\mu V_\mu S \rangle, \\
        \mathcal{L}_{SSV}&=& i g_{SSV}  \langle V^\mu [S, \partial_\mu S] \rangle, \label{Eq:L_SSV}
   \end{eqnarray}
where the symbol $\langle ... \rangle$ indicates the trace in the SU(3) flavor space.
The overall couplings $g_{SSS}$, $g_{SPP}$, $g_{SVV}$ have dimension of $[\text{mass}]$ and $g_{SSV}$ is a dimensionless constant.
The detailed forms of the flavor SU(3) multiplets can be found in~\cite{Cheng:2023lov}.

We use simple Breit-Wigner propagators for the $K^*$ and $\rho$, which are relatively narrow states. 
However, for the scalar mesons deep into the complex plane the Breit-Wigner parameterization is not accurate. 
Hence, as shown in Fig.~\ref{Fig:f01370decay},  the $\kappa$ and $\sigma$ appearing in the decay mechanism as intermediate states are represented as simple poles which are parameterized as
 \begin{eqnarray}
        G_{\kappa}(s)= \frac{1}{s- s_{\kappa}},~~ G_{\sigma}(s)= \frac{1}{s-s_{\sigma}}
 \end{eqnarray}
 respectively,
 where $s_{\kappa}$ and $s_{\sigma}$ are respectively the squares of the pole positions:
  $\sqrt{s_{\kappa}}=m_{\kappa}-i \Gamma_{\kappa}/2 $ with $m_{\kappa}=0.7$ GeV, $\Gamma_{\kappa}=0.6 $ GeV
 and $\sqrt{s_{\sigma}}= m_{\sigma}-i \Gamma_{\sigma}/2$ with $m_{\sigma}=0.47$ GeV, $\Gamma_{\sigma}=0.55$ GeV, the average values listed in the RPP~\cite{ParticleDataGroup:2024cfk}.

  It is important to note that several distinct
mechanisms contribute to the same final states, particularly in the $4 \pi$ channel. Consequently, the relative phases of the couplings must be explicitly taken into account to determine the interference pattern.
Expanding the Lagrangian in Eq.~(\ref{Eq:L_SPP}) and focusing on the $\sigma \pi \pi$ vertex,
we can obtain:
\begin{eqnarray}
        \mathcal{L}_{\sigma \pi \pi} =\frac{g_{SPP}}{\sqrt{2}} \cdot \sigma \pi^0 \pi^0 + \sqrt{2} g_{SPP} \cdot \sigma \pi^+ \pi^-.
\end{eqnarray}
Note that due to the identical feature,
there are two ways to contract the two $\pi^0$ fields and the two final $\pi^0$ states for the first term.
Based on this effective Lagrangian, the effective couplings are defined as follows: $g_{\sigma \pi^0 \pi^0}= g_{SPP}/\sqrt{2}$ and $g_{\sigma \pi^+ \pi^-} = \sqrt{2} g_{SPP}$.
The effective couplings $g_{\kappa K \pi }$ and $g_{\sigma \pi \pi}$ we define here can be related to the residues at the resonance poles of $\kappa$ in $K \pi$ scattering and $\sigma$ in $\pi \pi$ scattering,  respectively.
 We adopt the residue $\tilde{g}_{\sigma \pi \pi}= 3.61 e^{-1.30 i } $ GeV for the $\sigma$ pole from Refs.~\cite{Hoferichter:2023mgy,Garcia-Martin:2011nna}.
 By transforming this coupling for $\pi \pi$ with fixed isospin $I=0$ to the coupling associated with specific particles based on Eq.~(\ref{Eq:ff-sigma}), we have 
 $g_{\sigma \pi^+ \pi^-}=\tilde{g}_{\sigma \pi \pi}(-\sqrt{2}/\sqrt{3})=-0.79+2.84 i $ GeV and $g_{\sigma \pi^0 \pi^0}=g_{\sigma \pi^+ \pi^-}/2=-0.39+1.42 i$ GeV. 
 Ref.~\cite{Hoferichter:2023mgy} also provides the coupling $\tilde{g}_{\rho \pi \pi}=6.01 e ^{-0.09 i}$.
 
We adopt the residues $\tilde{g}_{\kappa^0 K\pi}= 3.81 e^{-1.49 i }$ GeV and $\tilde{g}_{K^{*0} K \pi}= 5.69 e ^{- 0.076 i}$ from Ref.~\cite{Pelaez:2020gnd}.
 Similarly, they can be transferred to the couplings associated with specific particles based on Eq.~(\ref{Eq:ff-kappa}): $g_{\kappa^0 K^+ \pi^-}=\tilde{g}_{\kappa^0 K\pi} (\sqrt{2}/{\sqrt{3}})= 0.25 - 3.10 i$ and 
$g_{K^{*0} K^+ \pi^-}=\tilde{g}_{K^{*0} K^+ \pi^- } (\sqrt{2}/{\sqrt{3}})= 4.63 -0.35 i$.
Because the pole positions of $\rho$ and $K^*$ are relatively close to the real axis,
their corresponding residues have small argument and are almost real numbers.
The strength of coupling $|g_{K^{*0} K^+ \pi^-}|=4.64$ is very close to the strength of $g_{\phi K^+ K^-}$ extracted from the decay of $\phi \to K \bar{K}$, e.g., $|g_{\phi K^+ K^-}|=4.47$ determined in Ref.~\cite{Cheng:2021nal}.
Such a phenomenon is a confirmation of the flavor symmetry result that $g_{K^{*0} K^+ \pi^-}= g_{\phi K^+ K^-}=g_{VPP}$.

Now we turn to the discussion of the remaining relevant couplings. 
The magnitude of $g_{\sigma \pi^+ \pi^- }$ adopted here is $2.95$.
Similarly, from the expanded Lagrangian we adopt that $g_{\sigma K\bar{K}}= g_{\sigma \pi^+ \pi^-} /2$ under the SU(3) symmetry,
leading to the ratio of $g_{\sigma K \bar{K}}/ g_{\sigma \pi^+ \pi^-}$ is $0.5$. 
These couplings are also extracted in Ref.~\cite{Kaminski:2009qg} by fitting the $\pi \pi \to \pi \pi / K \bar{K}$ scattering data with different models of $K$- and $S$-matrices.
They found the average value of $|g_{\sigma \pi^+ \pi^-}|=2.47 \pm 0.45$ GeV and about $0.6 \pm 0.2$ for the ratio of $g_{\sigma K \bar{K}}/ g_{\sigma \pi^+ \pi^-}$, both quantities are consistent with the our adopted values. 

The $\sigma \kappa \kappa $ coupling can be approximately related to the $\sigma K \bar{K}$ and $\sigma \pi \pi$ couplings
 when the component $\pi$ or $K$ of the $\kappa$ is considered as a spectator.
Analogous estimations can be applied to the coupling $g_{\sigma \kappa \bar{K}^*}$ which can be related to the couplings $g_{K^{*0}K^+ \pi^-}$,
and the coupling $\rho \kappa \bar{\kappa}$ which can be related to the couplings $\rho \pi \pi $ and $\rho K \bar{K}$.
The $\rho K^* \bar{\kappa}$ coupling can be related to another $VVS$ type coupling $\sigma \rho \rho$ under flavor symmetry with 
coupling $g_{\sigma \rho^+ \rho^-}$ determined by assuming $g_{\sigma \pi^+ \pi^-}= g_{\sigma \rho^+ \rho^-}$. 
For convenience, we collect the explicit formulae for estimating these couplings in Appendix~\ref{Appendix:B} 
and list their specific values in Table~\ref{Tab:couplings}.

Based on these effective Lagrangians, the amplitude of the diagram in Fig.~\ref{Fig:f01370decay}(a) decaying to $\pi^+ \pi^- $ via the neutral $\kappa ^0 \bar{\kappa}^0$ component is written as
\begin{eqnarray}
        i \mathcal{M}^{N}_{\pi^+\pi^-}= \int \frac{d^4 p_1}{(2 \pi)^4} 
         \frac{- g_{F \bar{\kappa}^0 \kappa^0} \cdot g_{\bar{\kappa}^0 K^- \pi^+}
          \cdot g_{\kappa^0 K^+ \pi^-}}{(p^2_1-s_{\kappa})(p^2_2- m^2_{K})(p^2_3-s_{\kappa})}\mathcal{F}(p^2_2,m^2_{K}) .  \label{amp:pipi}         
\end{eqnarray}

Similarly, we can obtain the amplitude in this mechanism via the charged $\kappa^+ \kappa^-$ component and denoted it as $\mathcal{M}^C_{\pi^+\pi^-}$.
Notice that in the $\pi^+ \pi^-$ channel no charge conjugated loop can contribute for both the neutral and charged $\kappa$ components.
The kinematic variables of the particles in the triangle loop are also denoted in Fig.~\ref{Fig:f01370decay}(a).
The effective Lagrangian approach, which deals with the nonlocal effects by introducing empirical form factors to cut off the divergences,
are applied to broader kinematic regions. Hence, we include a commonly adopted monopole form factor to the exchanged mesons~\cite{Cheng:2004ru},
\begin{eqnarray}
        \mathcal{F}(p^2_\text{ex}, m^2_\text{ex})= \frac{\Lambda^2- m^2_\text{ex}}{\Lambda^2-p^2_\text{ex}},
\end{eqnarray} 
where $\Lambda = m_{\text{ex}} + \alpha \Lambda_{\text{QCD}}$ with $\Lambda_{\text{QCD}}=300~\text{MeV}$,
and $\alpha$ is a dimensionless cut-off parameter. $p_{\text{ex}}$ and $m_{\text{ex}}$ are the four-momentum and mass of the exchanged meson, respectively.
   
 The physical meaning of the cut-off $\Lambda$ is that it reflects the finite spatial size of the exchanged hadron.
The monopole form factor with $\Lambda_{\text{QCD}}\sim 300$ MeV was devised for typical $q\bar{q}$ mesons,
whose size is governed by the quark-gluon confinement scale. For such narrow $q\bar{q}$ states ($\pi$, $K$, $\rho$, $K^*$, etc.),
we reasonably choose $\alpha = 1, 2, 3$ in our calculation to examine the cut-off dependence.

However, for the wide scalar $\kappa$ exchanged in the triangle loops, the physical picture is
different: as molecular or non-$q\bar{q}$ states, they are loosely bound systems of constituent hadrons rather
than compact quark-gluon bound states. Their spatial extension is governed by hadronic binding scales,
which are softer than the quark confinement scale. 
Consequently, a lower cut-off is physically appropriate
for these states. In the absence of a well-established form factor prescription for wide molecular states,
we adopt a simple prescription: for the exchanged $\kappa$, we shift the cut-off parameter
downward by roughly $0.5 \sim 1$ compared to that for ordinary $q\bar{q}$ mesons. Specifically,
we use $\alpha_{B} = 0.5, 1, 2$ as the cut-off parameters for the exchanged wide scalars, respectively corresponding to the $\alpha = 1, 2, 3$ for
the narrow $q\bar{q}$ states.
However, due to the lack of knowledge about the behavior of the counter terms, 
yet another model dependence will be present in association with the cut-off energies and different forms for the form factors~\cite{Guo:2010ak}.

Based on the flavor symmetry, we have $g_{\kappa^0 K^+ \pi^-} = g_{\kappa^+ K^0 \pi^+}= \sqrt{2} g_{\kappa^+ K^+ \pi^0}=- \sqrt{2} g_{\kappa^0 K^0 \pi^0} $.
In the isospin symmetry limit, the total partial width of the $\pi \pi$ channel is calculated by 
\begin{eqnarray}
        \Gamma_{\pi \pi} = \frac{3}{2} \times  \frac{1}{2 m_{F}} \int d \Phi_{\pi \pi} \bigg| \mathcal{M}^{N}_{\pi^+ \pi^-} + \mathcal{M}^{C}_{\pi^+ \pi^-}  \bigg|^2,
\end{eqnarray} 
where $\Phi_{\pi \pi}$ indicate the $\pi \pi$ phase space. The factor $3/2$ originates from the inclusion of the $\pi^0 \pi^0$ final state considering 
their identical feature and the relations in couplings.

The mechanism for the $f_0(1370)$ decay to $\eta \eta$ is the same as that for  decay to $\pi^+ \pi^-$, where just changing the vertices of $\kappa \to K \pi$ to $\kappa \to K \eta$. Hence, by replacing these couplings we can obtain the amplitudes $\mathcal{M}^C_{\eta \eta}$ and $\mathcal{M}^{N}_{\eta \eta}$  for the $\eta \eta$ final states. 
Note that the charge conjugated loop can contribute for both the neutral and charged $\kappa$ components in this channel.
The decay width of the $\eta \eta$ channel is given by 
\begin{eqnarray}
        \Gamma_{\eta \eta} = \frac{1}{2} \times  \frac{1}{2 m_{F}} \int d \Phi_{\eta \eta} \bigg| 2 \mathcal{M}^{N}_{\eta \eta} + 2 \mathcal{M}^{C}_{\eta \eta}  \bigg|^2,
\end{eqnarray} 
$\Phi_{\eta \eta}$ is the phase space of $\eta \eta $ final states and $1/2$ is the symmetry factor.

 \begin{table}
        \caption{Collections of the relevant couplings. The mixing angle $\theta= -10.7^\circ$ for $\eta$-$\eta'$ mixing in the singlet-octet basis is reasonably adopted~\cite{Cheng:2024sus}.}
        \begin{tabular}{l  c}
                \hline \hline 
                 Couplings  &  Flavor symmetry   \\
                 \hline
                 $g_{\kappa^0 K^+ \pi^-}= 0.25 - 3.10 i$ GeV~\cite{Pelaez:2020gnd} \qquad \qquad                                            &  $g_{SPP}=g_{\kappa^0 K^+ \pi^-} = \sqrt{2} g_{\kappa^+ K^+ \pi^0}= g_{\kappa^+ K^0 \pi^+}=- \sqrt{2} g_{\kappa^0 K^0 \pi^0} $ \\
                 $g_{\kappa^0 K^0\eta} = -0.10 +1.24 i $     GeV                                                                           &  $g_{\kappa^0 K^0 \eta}  = g_{SPP}\frac{-1}{\sqrt{6}} \cos\theta  $   \\

                $ g'_{\kappa^0  K^0 \eta}=-0.04 + 0. 58 i $  GeV &  $g'_{\kappa^0 K^0 \eta} =   g_{SPP}[\frac{-1}{\sqrt{6}} \cos\theta + \frac{2}{\sqrt{3}} (-\sin \theta )]$ \\
                 \hline 
                 $g_{\sigma \pi^+ \pi^- }= -0.79 + 2.84 i$ GeV ~\cite{Hoferichter:2023mgy}                                                 &   $2 g_{\sigma \pi^0 \pi^0}= g_{\sigma \pi^+ \pi^-}= 2 g_{\sigma K \bar{K}}$ \\
                 $g_{\sigma K^0 \bar{K}^0} = -0.39 + 1.42 i$ GeV                                                                           &   $g_{\sigma K^+ K^-} = g_{\sigma K^0 \bar{K}^0}$                \\
               $g_{\sigma \kappa^0 \bar{\kappa}^0 } =  3 g_{\sigma K^0 \bar{K}^0}=-1.18 +4.26 i $  GeV                                     &   $g_{\sigma \kappa^+ \kappa^-}= g_{\sigma \kappa^0 \bar{\kappa}^0}$  \\
                 $g_{\sigma \kappa^0 \bar{K}^{*0}} = 3.27 -0.25 i  $  GeV                                                                     &   $g_{\sigma \kappa^+ K^{*-}}=g_{\sigma \kappa^0 \bar{K}^{*0}}= g_{VPP} /\sqrt{2} $  \\
                 \hline
                 $g_{K^{*0}K^+ \pi^-}= 4.63 -0.35 i $   ~\cite{Pelaez:2020gnd}                                                           &   $g_{VPP}=g_{K^{*0}K^+ \pi^-}$        \\                    
                 $g_{\rho^0 \pi^+ \pi^-}= -5.98 + 0.54 i $ ~\cite{Hoferichter:2023mgy}                                                    &   $ \sqrt{2}g_{VPP}=g_{\rho^+ \pi^+ \pi^0}=-g_{\rho^- \pi^- \pi^0}=- g_{\rho^0 \pi^+ \pi^-}$          \\
                 $ g_{\rho^0 K^0 \bar{K}^0}=3.27 -0.25 i  $                                                                              &   $ g_{\rho^0 K^0 \bar{K}^0}=g_{VPP} /\sqrt{2}$     \\
                \hline 
                 $g_{\rho^0 \kappa^0 \bar{\kappa}^0 } = g_{\rho^0 K^0 \bar{K}^0}=3.27 -0.25 i  $                                          &   $g_{\rho^+ \kappa^+ \bar{\kappa}^0}= \sqrt{2} g_{\rho^0 \kappa^+ \kappa^-}$       \\
                 $g_{\sigma \rho^+ \rho^-} =g_{\sigma \pi^+ \pi^-}$                                                                       &   $g_{\sigma \rho^+ \rho^-}= \sqrt{2} g_{VVS}$                        \\
                  $g_{\rho^0 \kappa^+ K^{*-}}=g_{VVS}/ \sqrt{2} = -0.39 +1.42 i  $ GeV                                      &   $g_{\rho^+ \kappa^+ \bar{K}^{*0}}= \sqrt{2} g_{\rho^0 \kappa^+ \bar{K}^{*-}}$, \quad  $g_{\rho^0 \kappa^0 \bar{K}^{*0}}=-g_{\rho^0 \kappa^+ K^{*-}}$   \\
                 \hline
                 $g_{SPP}= 0.25 - 3.10 i$ GeV       &   -  \\
                 $g_{VPP}= 4.63 - 0.35 i$ GeV       &   -   \\
                 $g_{VVS}= -0.55 + 2.01 i$ GeV      &   - \\
                 \hline \hline 
        \end{tabular}
        \footnotetext{The relations in the flavor symmetry column are derived from SU(3) symmetry. The numerical values of the couplings on the left side are extracted 
        from different phenomenological analyses and therefore do not strictly satisfy these SU(3) relations. 
        For instance, $g_{\rho^0 \pi^+ \pi^-}$ from Ref.~\cite{Hoferichter:2023mgy} gives $g_{VPP}=4.23-0.38 i $, which slightly deviates from $g_{VPP}=g_{K^{*0}K^+\pi^-}$ from Ref.~\cite{Pelaez:2020gnd}.
        }
        \label{Tab:couplings}
\end{table}

 The coupling $g_{\kappa K \eta }$ can be related to $g_{\kappa^0 K^+ \pi^-}$ by the flavor symmetry.
 In fact, the physical $\eta$ state is the mixing of $\eta_8$ and $\eta_1$ with $\eta=\cos \theta  \eta_8 -\sin \theta \eta _1$ due to the SU(3) flavor breaking.
 As shown in Table~\ref{Tab:couplings}, we adopt two strategies to deal with the $g_{\kappa K \eta}$ coupling: one is that we ignore the $\eta_1$ component in $\eta$ by taking $\eta=\cos \theta \eta_8$, and restrict to the symmetry inside the octet by using $P_8$ as the pseudoscalar matrix. In this case, we denote this coupling as $g_{\kappa K \eta}$; the other case is that we take $\eta=\cos \theta  \eta_8 -\sin \theta \eta_1  $ and relate the couplings associated with these two part as well as the $g_{\kappa K \pi}$ by the U(3) flavor symmetry, where the pseudoscalar matrix is taken as $U$. In this case, this coupling is denoted as $g'_{\kappa K \eta}$. The explicit forms of $P_8$ and $U$ can be found in Appendix.~\ref{Appendix:C}.
 One must account for the fact that the $U(3)$ flavor symmetry is not good enough in the pseudoscalar sector due to the U(1) anomaly.

Similarly, as shown in Fig.~\ref{Fig:f01370decay}(b), the amplitude of $f_0(1370)$ decay to $K^+ K^-$ channel by exchanging $\pi$ via the neutral $\kappa^0 \bar{\kappa}^0$ component is written as 
\begin{eqnarray}
        i \mathcal{M}^{N}_{K^+ K^-}=\int \frac{d^4 p_1}{(2 \pi)^4}  \frac{ -g_{F \bar{\kappa}^0 \kappa^0} \cdot g_{\bar{\kappa}^0 K^- \pi^+} \cdot g_{\kappa^0 K^+ \pi^-} }{(p^2_1-s_{\kappa})(p^2_2- m^2_{\pi})(p^2_3-s_{\kappa})}\mathcal{F}(p^2_2,m^2_{\pi}).
\end{eqnarray} 
Same as the case in $\pi \pi$ channel, there is no charge conjugated loop that can contribute.
Considering the relation between the couplings $g_{ {\kappa^0} K^+ \pi^-}$ and $g_{ {\kappa^+} K^+ \pi^0}$, we find that $\mathcal{M}^{C}_{K^+K^-}=(1/2) \mathcal{M}^{N}_{K^+K^-}$.
The amplitude of $f_0(1370)$ decay to $K^+ K^-$ channel by exchanging $\eta$ is written as 
\begin{eqnarray}
        i \mathcal{M}'_{K^+ K^-}=\int \frac{d^4 p_1}{(2 \pi)^4}  \frac{ -g_{F \kappa^+ \kappa^-} \cdot g_{\kappa^+ K^+ \eta} \cdot g_{\kappa^- K^- \eta} }{(p^2_1-s_{\kappa})(p^2_2- m^2_{\eta})(p^2_3-s_{\kappa})}\mathcal{F}(p^2_2,m^2_{\eta}).
\end{eqnarray} 
Then we obtain the total decay width to $K\bar{K}$ via
\begin{eqnarray}
        \Gamma_{K \bar{K}} =2 \times  \frac{1}{2 m_{F}} \int d \Phi_{K \bar{K}} \bigg| \mathcal{M}^{N}_{K^{+} K^-} + \mathcal{M}^{C}_{K^{+} K^-} + \mathcal{M}'_{K^{+} K^-}\bigg|^2,
\end{eqnarray} 
   where the $\Phi_{K \bar{K}}$ indicate the $K \bar{K}$ phase space. The factor $2$ originates from the inclusion of $K^0 \bar{K}^0$ final state.

For the  $4 \pi$ decay channel, there are three different specific final states which are $2 \pi^+ 2 \pi^-$, $\pi^+ \pi^- 2 \pi^0$ and $4 \pi^0$ respectively.
As claimed by the analyses in experiments~\cite{Bettini:1966zz,CrystalBarrel:2001xud,CRYSTALBARREL:2001ldq}, 
the $4 \pi$ final states are mainly from the intermediate channels of $\sigma \sigma$ and $\rho \rho$. This mechanism is shown in Fig.~\ref{Fig:f01370decay}(c).
Specifically, for the decay of an isoscalar state $f_0$, the $2 \pi^+2 \pi^-$ channel is only from the $\rho^0 \rho^0$ and $\sigma \sigma $ intermediate channels;
the $\pi^+ \pi^- 2 \pi^0$ channel is only from the $\rho^+ \rho^-$ and $\sigma \sigma$ intermediate channels;
and the $4 \pi^0$ channel is only from the $\sigma \sigma$ intermediate channel. 
Because these final states include identical particles, we must account for additional contributions arising from identical particles exchange.

Firstly we focus on the decay width of the $2 \pi^+ 2 \pi^-$ channel, then the decay widths of the other two channels can be related to $\Gamma_{2 \pi^+ 2 \pi^-}$ by considering
the relations among the couplings in the isospin limit.
As shown in Fig.~\ref{Fig:f01370decay}(c), we consider the decay mechanism of $\kappa \bar{\kappa}$ rescattering to $\rho \rho$ or $\sigma \sigma$ first and subsequently decaying to $4\pi$.
The amplitude of the process for $\kappa \bar{\kappa}$ rescattering to $2\sigma\to 2 \pi^+ 2\pi^-$ by exchanging scalar $\kappa$, 
denoted as $\mathcal{M}^{S}_{\sigma}$, has the form of: 
\begin{eqnarray}
        i \mathcal{M}^{S}_{\sigma}= 4 \times \int \frac{d^4 p_1}{(2\pi)^4}
        \frac{- g_{F \kappa^0 \bar{\kappa}^0}\cdot g^2_{\sigma \kappa^0 \bar{\kappa}^0 } \cdot g^{2}_{\sigma \pi^+ \pi^-}
         \mathcal{F}(p^2_2, m^2_{\kappa})}{(p^2_1-s_{\kappa})(p^2_2-s_{\kappa})(p^2_3-s_{\kappa})(p^2_{ab}-s_{\sigma})(p^2_{cd}-s_{\sigma})} + ex.~ ,
\end{eqnarray}
momenta of the final $\pi$ are denoted as $\pi^+({p_a}) \pi^-(p_b) \pi^+ (p_c) \pi^-(p_d)$, and $p_{ab}=p_a+p_b$, $p_{cd}=p_c+p_d$ are respectively the momenta of the two intermediate $\sigma$.
 The $ex.$ term indicates the contribution from exchanging the identical particles, which is the same as the first term but only change the momenta of the two $\sigma$ to $p_{ad}=p_a+p_d$ and $p_{bc}=p_b+p_c$ respectively.
Note that, for such a decay mechanism, there are two charge conjugated loops for both the charged and neutral $\kappa \bar{\kappa}$ rescattering. By analyzing the relations of the couplings involved, we find that the amplitudes of these four diagrams are identical, resulting in a multiplicative factor of $4$.

Similarly, the amplitude of the process shown in Fig.~\ref{Fig:f01370decay}(c) for the $\kappa \bar{\kappa}$ rescattering to $2\sigma\to 2 \pi^+ 2\pi^-$ by exchanging vector $K^{*}$, 
 denoted as $\mathcal{M}^{V}_{\sigma}$, has the form of 
 \begin{eqnarray}
        i \mathcal{M}^{V}_{\sigma}= 4 \times  G
        \int \frac{d^4 p_1}{(2 \pi)^4} \frac{(p_{ab}+p_1)_\mu (p_{cd}+p_3)_{\nu}(g^{\mu \nu} -\frac{p^\mu_2 p^\nu_2}{m^2_{K^*}}) \mathcal{F}(p^2_2,m_{K^*})}
        {(p^2_1-s_{\kappa})(p^2_2-m^2_{K^*})(p^2_3-s_{\kappa})(p^2_{ab}-s_{\sigma})(p^2_{cd}-s_{\sigma})} + ex.~,
 \end{eqnarray} 
   here for brevity we use $G =g_{F \kappa \bar{\kappa}^0} \cdot g^2_{\sigma \kappa^0 \bar{K}^{*0}} \cdot g^2_{\sigma \pi^+ \pi^-}$.

   The amplitude of the process for the $\kappa \bar{\kappa}$ rescattering to $2\rho^0 \to 2 \pi^+ 2\pi^-$ by exchanging $\kappa$, 
 as shown in Fig.~\ref{Fig:f01370decay}(c), has the form of 
 \begin{eqnarray}
        i \mathcal{M}^{S}_{\rho}= 4 \times G
        \int \frac{d^4 p_1}{(2 \pi)^4} \frac{ (p_1+p_2)_\mu (p_3 -p_2)_\alpha (g^{\mu \nu}- \frac{p^\mu_{ab} p^\nu_{ab}}{m^2_{\rho}})(g^{\alpha \beta}- \frac{p^\alpha_{cd} p^\beta_{cd}}{m^2_{\rho}})(p_a-p_b)_\nu (p_c-p_d)_\beta \mathcal{F}(p^2_2,m_{\kappa}) }
        {(p^2_1-s_{\kappa})(p^2_2-s_{\kappa})(p^2_3-s_{\kappa})(p^2_{ab}-m^2_{\rho}+i m_{\rho} \Gamma_{\rho})(p^2_{cd}-m^2_{\rho}+ i m_{\rho } \Gamma_{\rho})  }+ex. ~,  \nonumber \\
         \end{eqnarray} 
here $G=g_{F \kappa \bar{\kappa}^0} \cdot g^2_{\rho^0 \kappa^0 \bar{\kappa}^0} \cdot g^2_{\rho^{0} \pi^+ \pi^-}$.

The amplitude of the process for the $\kappa \bar{\kappa}$ rescattering to $2\rho^0 \to 2 \pi^+ 2\pi^-$ by exchanging $K^{*}$, 
as shown in Fig.~\ref{Fig:f01370decay}(c), has the form of 
\begin{eqnarray}
        i \mathcal{M}^V_{\rho}=4 \times G \int \frac{d^4 p_1}{(2 \pi)^4}
        \frac{(g_{\mu \alpha}- \frac{p_{2\mu } p_{2 \alpha}}{m^2_{K^*}})(g^{\mu \nu}- \frac{p^\mu_{ab } p^\nu_{ab}}{m^2_{\rho}})(g^{\alpha \beta}- \frac{p^\alpha_{cd} p^{\beta}_{cd}}{m^2_{\rho}}) (p_a-p_b)_{\nu}(p_c-p_d)_{\beta} \mathcal{F}(p^2_2,m_{K^*}) }
        {(p^2_1-s_{\kappa})(p^2_2-m^2_{K^*})(p^2_3-s_{\kappa})(p^2_{ab}-m^2_{\rho}+i m_{\rho} \Gamma_{\rho})(p^2_{cd}-m^2_{\rho}+i m_{\rho } \Gamma_{\rho})} + ex.~,
\end{eqnarray}
here $G=g_{F \kappa \bar{\kappa}^0} \cdot g^2_{\rho^0 \kappa^0 \bar{K}^{*0} }\cdot g^2_{\rho^0 \pi^+ \pi^-} $.
The total amplitude of decaying to $2 \pi^+ 2 \pi^-$ is the sum of all the amplitudes above, i.e.,
\begin{eqnarray}
        \mathcal{M}_{2 \pi^+ 2 \pi^-}=\mathcal{M}^{S}_{\sigma} + \mathcal{M}^{V}_{\sigma} + \mathcal{M}^{S}_{\rho} + \mathcal{M}^{V}_{\rho}.
\end{eqnarray}
The decay width of this final state is 
\begin{eqnarray}
        \Gamma_{2 \pi^+ 2\pi^-} = \frac{1}{4} \cdot \frac{1}{2 m_{F}}   \int d\Phi_{2 \pi^+ 2\pi^-} |\mathcal{M}_{2 \pi^+ 2 \pi^-}|^2.
\end{eqnarray}
The constant $1/4$ is the symmetry factor on account of the two kinds of identical particles: two $\pi^+ $ and two $\pi^-$.

Then, we can discuss the decay amplitudes of $\pi^+ \pi^- 2 \pi^0$ and $4 \pi^0$ states based on the amplitude of the $2 \pi^+ 2\pi^-$ state.
For the $\rho^+ \rho^- \to \pi^+ (p_a)\pi^0 (p_b) $-$\pi^- (p_c) \pi^0 (p_d)$ process, we find the following similarities and differences compared to the $\rho^0 \rho^0 \to \pi^+ (p_a)\pi^- (p_b) $-$\pi^+ (p_c) \pi^- (p_d)$:

\begin{itemize}
        \item There is no charge conjugated loop for both the charged and neutral $\kappa \bar{\kappa}$ rescattering.
        \item Comparison of the relations between couplings $g_{\rho^+ \kappa \bar{\kappa} } / g_{\rho^+ \kappa \bar{K^*}}$ and $g_{\rho^0 \kappa \bar{\kappa} }/g_{\rho^0 \kappa \bar{K^*}}$ shown in Table~\ref{Tab:couplings}.
        \item Including the diagram obtained from exchanging the two identical $\pi^0$.
        \item The relation of the coupling $g_{\rho^{\pm}\pi^{\pm}\pi^{0}}$ and $g_{\rho^{0} \pi^+ \pi^-}$ shown in Table~\ref{Tab:couplings}.
\end{itemize}
 Considering the above items we find that the amplitudes adapted to the $\rho^+ \rho^- \to \pi^+ \pi^- 2 \pi^0$ process are just $-\mathcal{M}^{S}_{\rho}$ and $- \mathcal{M}^{V}_{\rho}$.

For the $\sigma \sigma \to \pi^+ (p_a) \pi^- (p_c)$-$ \pi^0(p_b) \pi^0 (p_d)$ process, we find the following discrepancies compared to the $\sigma \sigma \to \pi^+ (p_a)\pi^- (p_b) $-$\pi^+ (p_c) \pi^- (p_d)$:
\begin{itemize}
    \item The exchange of the two identical $\pi^0$ from the same $\sigma$ is topologically identical.
    \item  The relation of couplings: $g_{\sigma \pi^+ \pi^-}=2 g_{\sigma \pi^0 \pi^0}$. 
    \item There are two ways to contract the two $\pi^0$ fields and the two final $\pi^0$ states at the $\sigma \pi^0 \pi^0$ vertex.
\end{itemize} 
Therefore, the amplitudes $\mathcal{M}^S_{\sigma}$ and $\mathcal{M}^V_{\sigma}$ adapted to the $\pi^+ \pi^- 2 \pi^0$ state 
 are obtained by changing the momenta of the two $\sigma$ to $p_{ac}=p_a+p_c$ and $p_{bd}=p_b+p_d$ respectively and eliminating the exchange terms.

The decay width of the $\pi^+ \pi^- 2 \pi^0$ state is 
\begin{eqnarray}
        \Gamma_{ \pi^+ \pi^- 2 \pi^0} = \frac{1}{2} \cdot \frac{1}{2 m_{F}}   \int d\Phi_{\pi^+ \pi^- 2 \pi^0} |\mathcal{M}_{\pi^+ \pi^- 2 \pi^0}|^2.
\end{eqnarray}
The constant $1/2$ is the symmetry factor on account of the two identical particles $\pi^0 $.

For the $\sigma \sigma \to 4 \pi^0$ process, the amplitudes $\mathcal{M}^S_{\sigma}$ and $\mathcal{M}^V_{\sigma}$ for $2\pi^+ 2\pi^-$ could be directly adopted but need to further considering
all possible exchanges of the $\pi^0$.
The decay width of the $4 \pi^0$ state is 
\begin{eqnarray}
        \Gamma_{ 4 \pi^0} = \frac{1}{4!} \cdot \frac{1}{2 m_{F}}   \int d\Phi_{4 \pi^0} |\mathcal{M}_{4 \pi^0}|^2.
\end{eqnarray}
The constant $1/4!$ is the symmetry factor on account of the four identical particles $\pi^0 $.

For the $K \bar{K} \pi \pi $ channel from the tree-level decay $f_0(1370) \to \kappa \bar{\kappa} \to (K \pi ) (\bar{K} \pi)$ shown in Fig.~\ref{Fig:f01370decay}(d), 
there are six different final states specifically. The $K^+ K^- \pi^+ \pi^-$ and $K^0 \bar{K}^0 \pi^0 \pi^0$ can only originate from the $\kappa^0 \bar{\kappa}^0$ component;
the $K^0 \bar{K}^0 \pi^+ \pi^-$ and  $K^+ K^- \pi^0 \pi^0$ can only originate from the $\kappa^+ \kappa^-$ component;  
while the $K^0 K^- \pi^+ \pi^0$ and $K^+ \bar{K}^0 \pi^0 \pi^-$ can originate from both the $\kappa^0 \bar{\kappa}^0$ and $\kappa^+ \kappa^-$ components.

For the $K^+ K^- \pi^+ \pi^+$ final states, we can write down the amplitude as following:
\begin{eqnarray}
        \mathcal{M}_{K^+ K^- \pi^+ \pi^-}= g_{F \kappa^0 \bar{\kappa}^0 } \cdot g^2_{\kappa^0 K^+ \pi^-}  G_{\kappa}(s_{ab}) G_{\bar{\kappa}}(s_{cd}) ,
\end{eqnarray}
where $s_{ab}$ and $s_{cd}$ are respectively the invariant mass squares of the $K \pi$ pairs coming from $\kappa$ and $\bar{\kappa}$. 
The amplitudes of other final states can be obtained by replacing the couplings in $\mathcal{M}_{K^+ K^- \pi^+ \pi^-}$ and carefully including all the possible diagrams.
The combinations of $K \pi $ are definite for all the $K \bar{K} \pi \pi$ channels except for $K^+ K^- \pi^0 \pi^0$ and $K^0 \bar{K}^0 \pi^0 \pi^0$ due to the two identical particles $\pi^0$.
Hence, for these two final states, we need to include another diagram obtained by exchanging these two $\pi^0$ and when calculating the decay width, the symmetry factor $1/2$ should be considered on account of the double counting in the phase space.

Moreover, we address the implications of the SU(3) flavor symmetry for the scalar-scalar molecular picture adopted here.
If $f_0(1370)$ is a $\kappa\bar{\kappa}$ molecule, the flavor symmetry seems to imply the existence of a full molecular multiplet.
To study this issue, we have performed a systematic analysis of the vector-exchange induced effective potentials between scalar-meson pairs based on the $\mathcal{L}_{SSV}$ Lagrangian, Eq.~(\ref{Eq:L_SSV}).
For a given channel, the $t$- and $u$-channel potentials are evaluated at threshold via $V_{\text{eff}} = \sum_V g_{a}^{(V)} g_{b}^{(V)}/m_V^2$,
where $g_{a}^{(V)}$ and $g_{b}^{(V)}$ are respectively the couplings of the upper and lower vertices that exchange vector $V$. 
Then the effective potential $V^{I}_{\text{eff}}$ with specific isospin is obtained by projecting the total potentials onto states of fixed total isospin using the convention adopted here.
 
For simplicity, we set the overall coupling $g_{SSV}=1$, the quantity $V^{I}_{\text{eff}}$ should therefore be interpreted as relative indicators for comparing attraction or repulsion strengths across different channels, 
rather than as absolute potential strengths.
The resulting effective potentials (in units of GeV$^{-2}$) are summarized in Table~\ref{Tab:SSV_potentials}.
  
\begin{table}[htbp]
\centering
\caption{Effective potentials $V^{I}_{\text{eff}}$ for scalar-meson pair channels from $t$- and $u$-channel vector-meson exchange, evaluated with $g_{SSV}=1$.}
\label{Tab:SSV_potentials}
\begin{tabular}{ccl}
\hline \hline
Channel \qquad                              &  \qquad Isospin \qquad \qquad  & $V^{I}_{\text{eff}}$ (GeV$^{-2}$) \\
\hline
$\kappa\bar{\kappa}$                        & $I=0$                   & $-4.27$ (Strongly attractive) \\
$\kappa\bar{\kappa}$                        & $I=1$                   & $-0.95$ (Weakly attractive)   \\
\hline
$a_0 a_0$                                   & $I=0$                   & $-13.3$ (Strongly attractive) \\
$a_0 a_0$                                   & $I=1$                   & $0$ (Exact cancellation)      \\
$a_0 a_0$                                   & $I=2$                   & $+6.66$ (Repulsive)           \\
\hline 
$\sigma\sigma$, $\sigma a_0$, $\sigma f_0$, $f_0 f_0$, $a_0 f_0$      & $I=0 /1$                  & $0$ (Forbidden)    \\
\hline
$a_0\kappa$                                 & $I=1/2$                 & $-3.19$ (Attractive)          \\
$a_0\kappa$                                 & $I=3/2$                 & $+2.92$ (Repulsive)           \\
\hline
$f_0\kappa$                                 & $I=1/2$                 & $+1.26$ (Repulsive)           \\
$\sigma\kappa$                              & $I=1/2$                 & $+0.63$ (Repulsive)           \\
\hline \hline
\end{tabular}
\end{table}

The $\kappa\bar{\kappa}$ channel in $I=0$ receives attractive contributions from $\omega$, $\rho^0$, $\phi$, and $\rho^\pm$ exchange, providing independent support for the $\kappa\bar{\kappa}$ molecular interpretation of $f_0(1370)$.
The vanishing of $a_0 a_0$ in $I=1$ follows exactly from the antisymmetry of the scattering amplitude.
Among the mixed channels, only $a_0\kappa$ in $I=1/2$ is attractive, driven by off-diagonal $\rho^\pm$ exchange.
It is noteworthy that this pattern of attraction is analogous to the pseudoscalar-meson sector, 
where scalar resonances are only observed in $\pi\pi$ ($I=0$), $K \pi$ ($I=1/2$), and $K\bar{K}$ ($I=0,1$).
The $a_0 a_0$ $I=0$ molecule might be assigned to the $f_0(2020)/f_0(2100)/ f_0(2200) $ listed in the RPP, and the
$ a_0 \kappa$ $I=1/2$ molecule might be assigned to the $K^*_0(1430) / K^*_0(1950)$.
We note that this analysis is restricted to vector-meson exchange,
which is expected to be the dominant mechanism; other possible
contributions are not considered here.

\section{Numerical results and discussions} \label{Sec:results}

\begin{figure}
        \centering
       \includegraphics[width=3.0in ]{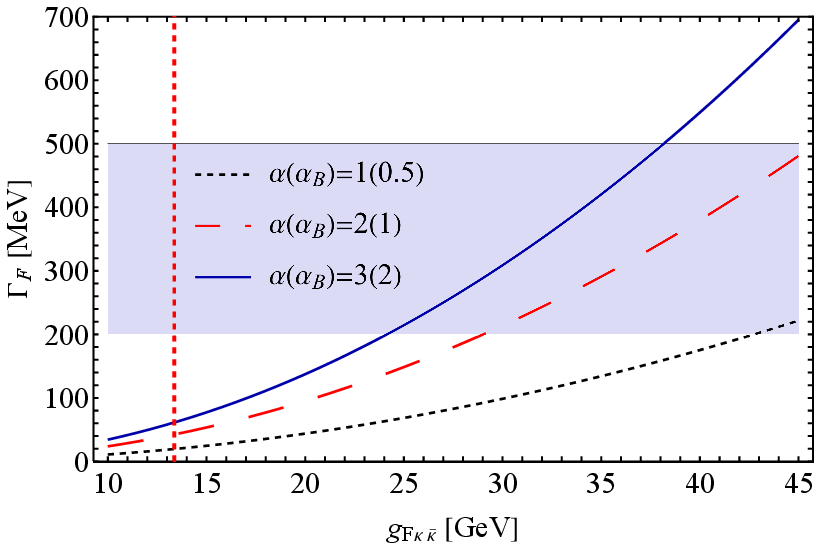}
      \caption{The total width of $f_0(1370)$ dependent on the parameter $g_{F \kappa \bar{\kappa}}$ with $\sqrt{s}=1.37$ GeV and three form factor parameter sets $\alpha (\alpha_B)= 1(0.5)/ 2 (1) /3 (2)$.
       The red dashed vertical line indicates the reference value $g_{F\kappa \bar{\kappa}}=13.4$~GeV which is estimated from the Weinberg compositeness criterion~\cite{Weinberg:1962hj,Weinberg:1965zz} using a complex $\kappa$~mass $\sqrt{s_{\kappa}}=0.7-0.3i$~GeV with $X=1$; 
      The light purple band indicates the range of the Breit-Wigner width of $f_0(1370)$ measured in experiments.} \label{Fig:f01370width}
\end{figure}

The partial decay widths of different channels we calculated including $K \bar{K}$, $\pi \pi$, $\eta \eta$, $4 \pi$ and $K\bar{K} \pi \pi $,
These are the primary strong decay channels of $f_0(1370)$, 
which are approximately summed up to the total width of $f_0(1370)$: the $\Gamma_F$.
We first present the total width at $\sqrt{s}=1.37$ GeV with $g_{F\kappa\bar{\kappa}}$ treated as a free parameter in Fig~\ref{Fig:f01370width}.
We take this couplings as a parameter to fit the total width of $f_0(1370)$ measured in experiments. 
 We will present results for two different couplings $g_{\kappa K \eta}$ and $g'_{\kappa K \eta}$ in this section. 
 For clarity, we denote the result associated with $g_{\kappa K \eta}$ as $\Gamma_F$ and with $g'_{\kappa K \eta}$ as $\Gamma'_F$.
We approximate the total $4\pi$ width by summing the $\sigma\sigma$ and $\rho\rho$ contributions, under the assumption that other intermediate states are negligible.
In our framework, the $f_0(1370)$ decaying to the $K \bar{K}, \pi \pi, \eta \eta$ and $4 \pi$ channels are via the triangle loop diagrams,
 which are dependent on the cut-off parameters $\alpha$. Hence, we also present the results with different cut-off parameters $\alpha(\alpha_B)$ from $1(0.5) \sim 3(2)$ in Fig~\ref{Fig:f01370width},
 to study the cut-off dependence.
The light purple band in Fig.~\ref{Fig:f01370width} indicates the range of the Breit-Wigner width of $f_0(1370)$ measured in experiments.  
 The pole width $\Gamma = -2\, \mathrm{Im}\, \sqrt{s_{\text{pole}}}$ for $f_0(1370)$ is estimated to be around $500$ MeV according to its pole position,
and its Breit-Wigner width is measured to be between $200$ MeV and $500$ MeV depending on the models and the reactions. 
The red dashed vertical line indicates the reference value $g_{F\kappa \bar{\kappa}}=13.4$~GeV, as estimated from the Weinberg criterion~\cite{Weinberg:1962hj,Weinberg:1965zz}
using a complex $\kappa$~mass $\sqrt{s_{\kappa}}=0.7-0.3i$~GeV and compositeness $X=1$.
As shown in Fig~\ref{Fig:f01370width}, with the $g_{F \kappa \bar{\kappa}}$ determined by the Weinberg criterion, the total width $\Gamma_F$
is much smaller than the measured values of $f_0(1370)$ collected in the RPP~\cite{ParticleDataGroup:2024cfk}.
  It is expected that the Weinberg criterion does not work for our broad component case even if a complex mass is used because it is derived from the situation of weakly bound state consisting of two stable or nearly stable components.

From Fig.~\ref{Fig:f01370width} we can see that when cut-off parameter $\alpha(\alpha_B)=1(0.5)$, the coupling $g_{F\kappa \bar{\kappa}}$ is taken as large as $45$ GeV,
the total width $\Gamma_F$ could merely reach the lower limit of the experimental width range.
While with $\alpha(\alpha_B)=2(1)$, $\Gamma_F$ can reach the experimental observation when $g_{F\kappa \bar{\kappa}}$ is around $30$ GeV.
 Certainly with $\alpha(\alpha_B)=3(2)$, the required $g_{F\kappa \bar{\kappa}}$ to reproduce the experimental width is further reduced to about $25$ GeV.
For presenting the accurate quantities of the partial widths of different channels when the total width is fitted to experiments,
we show the results in Table~\ref{Tab:Partialdecay} with $g_{F\kappa \bar{\kappa}}=35$ GeV.
 Notice that $g_{F\kappa \bar{\kappa}}$ is a common coupling for all the decay channels, its change will
 not affect the ratios among the decay channels.
  In other words, the decay pattern is independent of this coupling.

 From Table~\ref{Tab:Partialdecay}, we can see that when $\alpha(\alpha_B)$ varies from $2(1)$ to $3(2)$,
 the decay widths are relatively stable.
These decay channels have been observed in experiments except the $K\bar{K}\pi \pi$ channel, which lacks measurements up to now. 
At the nominal mass of $f_0(1370)$: $\sqrt{s}=1.37$ GeV, the $K\bar{K}$ channel has the largest partial width, followed by the $4 \pi$ and $\pi \pi $ channels.
The $\pi \pi$ and $4 \pi $ channels have comparable partial widths.  We can find that the partial width of $K\bar{K}$ channel is not sensitive to the strategy that determining the $\kappa K \eta $ coupling because the most dominant contribution to this channel is the mechanism of exchanging $\pi$. 

From Table~\ref{Tab:couplings}, we find that the U(3) symmetry expression of the $g'_{\kappa K \eta}$ coupling consists of two destructive terms,
which respectively correspond to the $\eta_8$ and $\eta_1$ parts in the mixing wave function. We adopt a relatively large value of $\theta=-10.7^\circ$ referring to the analysis in~\cite{Cheng:2021nal, Cheng:2024sus}.
Due to the comparable size of these two parts, the destructive interference leads to a small value of $g'_{\kappa K \eta}$, which is sensitive to the mixing angle $\theta$. 
Hence, the partial width $\Gamma'(F\to \eta \eta)$ calculated with $g'_{\kappa K \eta }$ is the smallest, with its order of magnitude being one-thousandth of the partial widths of the other channels. Compared to it, $\Gamma(F \to \eta \eta)$ is larger by a factor of about $20$ because the coupling $g_{\kappa K \eta}$ is not suppressed by the destructive interference.  
But $g_{\kappa K \eta}$ is still smaller than $g_{\kappa K \pi}$, which is contrary to the relation between $g_{K^* K \pi}$ and $g_{K^* K \eta}$: $g_{K^*K \pi ^ {\pm}}:g_{K^*K \eta}= 1: (\sqrt{3}/\sqrt{2}) \cos \theta $. Because the irreducible decomposition of $8 \otimes8$ include two different octet $8_A$ and $8_S$. Here  $A$ and $S$ indicate the antisymmetric and symmetric features corresponding to the $K^*$ and $\kappa$ flavor wave functions respectively, which leads to different coupling relations. Therefore, the calculations treating  the $f_0(1710)$ as a $K^* \bar{K}^*$ molecular state in Ref.~\cite{Wang:2021jub} show that its decays into $\eta \eta$ and $\pi \pi$ channels are in the same order of magnitude. 
While with the same triangle diagram mechanism exchanging $K$, the partial widths of $f_0(1370)$ treated as a $\kappa \bar{\kappa}$ molecule to these two channels differ by one to two orders of magnitude.

The $K\bar{K} \pi \pi$ channel is the only one proceeding through tree-level decay. However,
 the partial width of this channel is much smaller than that of the $K \bar{K}$, $\pi \pi $ and $4 \pi$ channels, which arise from loop-level decays. 
This phenomenon might be mainly attributed to the very limited $K \bar{K} \pi \pi$ phase space compared to the other channels around $\sqrt{s}=1.37$ GeV, 
which is near the $\kappa \bar{\kappa}$ threshold.
Consequently, around this energy region, the $K \bar{K} \pi \pi$ width is sensitive to the center-of-mass energy $\sqrt{s}$.

\begin{table}
        \caption{The partial widths of different channels in $f_0(1370)$ decay at fixed  $\sqrt{s} =1.37$ GeV,
               with $g_{F\kappa \bar{\kappa}}= 35.0$ GeV.
              For testing the cut-off dependence, the cut-off parameter for the loop diagrams is chosen as $\alpha(\alpha_B)=1(0.5)/\mathbf{2(1)}/3(2)$ for $K^* (\kappa)$exchange.
              $\Gamma_F$ is the sum of all the partial decay widths listed here, which can be approximately taken as the total width of $f_0(1370)$.
              We present results for two different couplings $g_{\kappa K \eta}$ and $g'_{\kappa K \eta}$ and denote the result associated with $g'_{\kappa K \eta}$ as $\Gamma'$.
               The decay widths are presented in unit of MeV.}
        \begin{tabular}{l | c  }
                \hline  \hline
                Decay channel                                                    &   \qquad     $g_{F\kappa \kappa}= 35.0$ GeV        \qquad                   \\
                 \hline
                  $\Gamma(F\to K \bar{K}) $          &    $47.5 / \mathbf{132} /207$                        \\

                  $\Gamma'(F \to K \bar{K})$   &    $43.5 / \mathbf{121} /189$                        \\ \hline
                   $\Gamma(F \to \pi \pi )   $                                    &    $22.1 / \mathbf{62.7} /103$                       \\  \hline
                   $ \Gamma(F\to \eta \eta )$                                     &    $0.72/  \mathbf{1.94} /3.08$                     \\
                   $ \Gamma'(F\to \eta \eta )$                                     &    $3.3 \times 10^{-2} /  \mathbf{9 \times 10^{-2}} /0.14$                     \\        \hline
                   $\Gamma(F \to  4 \pi)    $                                     &    $61.7 / \mathbf{92.6} /105.0$                          \\
                   \hline
                   $\Gamma( F \to  K \bar{K} \pi \pi) $                           &   $\mathbf{0.33}$  \\
                  \hline
                    $ \Gamma_{F}$                                                 &      $134 / \mathbf{291} /420$     \\
                     $ \Gamma'_{F}$                                                 &      $129 / \mathbf{278} /400$     \\   
                  \hline \hline
    \end{tabular}\label{Tab:Partialdecay}
\end{table}

\begin{table}
        \caption{Decay widths of $f_0(1370) \to 4\pi$ with $g_{F \kappa \bar{\kappa}}=35$ GeV. Exclusive widths for each diagram and different intermediate channels are also presented. 
                The notation [ex, $R \to ff$] in the second column specifies the exchange particle ``ex'' between the $\kappa$ pair and the intermediate resonance ``$R$'' decaying to final states ``$ff$''. 
                The decay widths are given in three different choices of the cut-off parameters $\alpha(\alpha_B)=1(0.5)/2(1)/3(2)$. Decay widths are evaluated at center-of-mass energy $\sqrt{s}=1.37$~GeV and given in unit of MeV.}
                \begin{tabular}{ l | c  | l |  c }
            \hline \hline  
             Channels                                & Diagram                                                      &     $\sqrt{s}=1.37$ GeV   \qquad                                &   The relevant couplings   \\
               \hline    
              \multirow{5}{*}{$2 \pi^+ 2 \pi^-$}     & $[\kappa,~\sigma \to \pi^+ \pi^- ]$                           &      $\Gamma_1 =6.5/ 8.5 / 14.6  $                                        &  $g_{\sigma \kappa \bar{\kappa} }$ , $g_{\sigma \pi^+  \pi^-}$  \\      
            
                                                     & $[K^*,~\sigma \to \pi^+ \pi^-  ] $                            &      $\Gamma_2 = 16.7/ 47.6/ 101  $                                        &  $g_{\sigma  \kappa \bar{K}^*}$,  $g_{\sigma \pi^+ \pi^-}$      \\
                                                  
                                                     & $\sigma \sigma \to  2 \pi^+ 2 \pi^-$                         &      $ \mathbf{30.9} / \mathbf{46.1} / \mathbf{49.9} $                                                         &                                                                 \\
                                                     \cline{2-4}
                                                     & $[\kappa,~\rho^0 \to \pi^+ \pi^-  ]$                         &      $\Gamma_3 =0.24/ 1.00/ 3.86 $                                         &   $g_{\rho^0 \kappa \bar{\kappa}}$,  $ g_{\rho^0 \pi^+ \pi^-}$    \\
                                                     & $[K^*, ~\rho^0 \to \pi^+ \pi^- ]$                            &      $\Gamma_4 = 0.10 /0.29 / 0.52 $                                          &   $f_{\rho^0 \kappa \bar{K}^*}$, $g_{\rho^0 \pi^+ \pi^-}$          \\
                                                     &  $\rho \rho \to 2 \pi^+ 2\pi^-$                              &      $ \mathbf{0.33} / \mathbf{1.20} / \mathbf{3.97} $                                                         &                                                                  \\
                                                     \cline{2-4}
                                                     & Total                                                        &     $ \mathbf{31.6} / \mathbf{47.4} / \mathbf{52.6} $                               &                                                                               \\
                                                     \hline \hline            
\multirow{5}{*}{$\pi^+ \pi^- 2 \pi^0$  }     &   $[\kappa, ~ \sigma \to \pi^+ \pi^- / \sigma \to \pi^0 \pi^0]$     &   $3.98 / 5.18 / 8.90  $                      &  $g_{\sigma \kappa \kappa}$, $g_{\sigma \pi^+ \pi^-}$, $g_{\sigma \pi^0 \pi^0}$  \\
                                             &   $[K^* ,~ \sigma \to \pi^+ \pi^- /\sigma \to \pi^0 \pi^0]$         &   $ 9.67/ 27.8/59.7 $                       &  $g_{\sigma \kappa \bar{K}^*}$ , $g_{\sigma \pi^+ \pi^-}$, $g_{\sigma \pi^0 \pi^0}$ \\
                                             &   $\sigma \sigma \to \pi^+ \pi^- 2 \pi^0$                           &   $ \mathbf{18.0}/ \mathbf{26.1} /\mathbf{27.9} $                           & \\
                                            \cline{2-4}
                                             &   $[\kappa,~ \rho^+ \to \pi^+ \pi^0 / \rho^- \to \pi^- \pi^0]$      & $ 2 \Gamma_3$                    &   $g_{\rho^{\pm} \kappa \bar{\kappa}}$, $g_{\rho^{\pm} \pi^{\pm} \pi^0}$ \\
                                             &   $[K^*,~ \rho^+ \to \pi^+  \pi^0  / \rho^- \to \pi^- \pi^0]$       & $ 2 \Gamma_4$                    &  $g_{\rho^{\pm} \kappa \bar{K}^*}$, $g_{\rho^{\pm} \pi^{\pm} \pi^0}$ \\   
                                             &   $\rho \rho \to \pi^+ \pi^- 2 \pi^0$                               & $ \mathbf{0.65}/ \mathbf{2.40}/ \mathbf{7.94}  $                             & \\
                                             \cline{2-4}
                                             &   Total                                                             & $\mathbf{18.9} / \mathbf{28.3} / \mathbf{34.1}  $                             &  \\
                                           \hline \hline
                       \multirow{3}{*}{$4 \pi^0$}          &     $[\kappa, ~ \sigma \to \pi^0 \pi^0]$               &  $ 2.26/ 2.95/ 5.09 $                       &  $g_{\sigma \kappa \bar{\kappa}}$, $g_{\sigma \pi^0 \pi^0}$ \\                                                 
                                                           &     $[K^*, ~ \sigma \to \pi^0 \pi^0] $                 &  $ 5.89/ 16.9/ 35.9 $                       &  $g_{\sigma \kappa \bar{K}^*}$, $g_{\sigma \pi^0 \pi^0}$\\
                                            \cline{2-4}
                                             &     Total                                                           &    $\mathbf{11.3}/\mathbf{16.9}/ \mathbf{18.4} $                            &   \\
                                            \cline{1-4}
                                             \hline \hline 
       \end{tabular} \label{Tab:4pi}
\end{table}


 With $g_{F\kappa \bar{\kappa} }=35$ GeV,
the partial widths of different specific $K\bar{K}\pi \pi$ final states at $\sqrt{s}=1.37$ GeV are as follows:  
$\Gamma(K^+ K^- \pi^+ \pi^-)= \Gamma(K^0 \bar{K}^0 \pi^+ \pi^-)=108 $ keV;
 $\Gamma(K^+K^- \pi^0 \pi^0 )=\Gamma(K^0 \bar{K}^0 \pi^0 \pi^0)=54 $ keV;
 $\Gamma(K^+ \bar{K}^0 \pi^0 \pi^-)=\Gamma(K^0 K^- \pi^+ \pi^0)=0.53 $ keV.
For the states with two charged $\pi$, the charged couplings $g_{\kappa K \pi^{\pm}}$ is larger than the neutral couplings by a factor of $\sqrt{2}$.
   While for the states including $\pi^0 \pi^0$, there are another exchanged diagram and an extra symmetry factor $1/2$, finally leading to such a relation.
   For the states including only one charged pion, there are two diagrams respectively from the two different $\kappa \bar{\kappa}$ components which interfere destructively due to the $SU(3)$ relations listed in Table~\ref{Tab:couplings}, 
   resulting in a very small partial width for these states. Such a pattern is different from the direct decay to $K\bar{K} \pi \pi$, such as the data in $J/\psi$ decay~\cite{ParticleDataGroup:2024cfk}.
  The same decay pattern among these three kinds of $K \bar{K} \pi \pi$ states should also appear in the $K^* \bar{K}^* \to (K \pi) (\bar{K}\pi)$ process due to the same relations of couplings in $K^*K \pi$ and $\kappa K \pi$.

  It is worth mentioning that in our framework, the decay widths to $K \bar{K}$, $\pi \pi$, $\eta \eta$ and $K \bar{K } \pi \pi$ are sensitive to the coupling $g_{\kappa K \pi}$ because they are essentially proportional to the fourth power of this coupling.
In fact, the $|\tilde{g}_{\kappa K \pi}|$ is determined to be $4.4$~\cite{Pelaez:2016klv} and $3.81$~\cite{Pelaez:2020gnd} in two different works respectively,
 while such a small difference in this coupling can lead to a variation in these partial widths by nearly a factor of two.

  The partial widths of different specific $4 \pi$ final states at $\sqrt{s}=1.37$ GeV are shown in Table~\ref{Tab:4pi}, 
 where we also present the exclusive contributions from different diagrams and intermediate states.
 The ratios between different $4 \pi$ states are closely related to the decay mechanisms, as the intermediate states $\rho \rho$ and $\sigma \sigma$ have
distinct features in $4 \pi$ decay. These differences can reflect the decay dynamics of the $f_0(1370)$ and thereby reveal its structure. 
Reviewing the respective amplitudes for $2 \pi^+ 2\pi^-$ and $\pi^+ \pi^- 2 \pi^0$ as well as the results at $\sqrt{s}=1.37$ GeV presented in Table~\ref{Tab:4pi},
we find that in our framework, the contributions from the $\sigma \sigma \to 2\pi^+ 2\pi^-$ are much larger than those from $\rho^0 \rho^0 \to 2 \pi^+ 2\pi^-$.
The widths of the process $\sigma \sigma \to \pi^+ \pi^- 2 \pi^0$ is smaller than that of $\sigma \sigma \to 2 \pi^+ 2 \pi^-$ even though the symmetry factor for $ \pi^+ \pi^- 2 \pi^0$ is larger.
This is mainly because there is no identical particle exchange diagram in the $\sigma \sigma \to \pi^+ \pi^- 2 \pi^0$ process where the two $\pi^0$ are from the same $\sigma$. 
In contrast, the process $\sigma \sigma \to 2 \pi^+ 2 \pi^- $ does include the contributions from such an exchange diagram. 
After the analysis it is found that the amplitudes for $\rho \rho$ decay to $2\pi^+ 2\pi^- $ and $\pi^+ \pi^- 2\pi^0$ are the same then their widths only differ by the symmetry factor, which bring the relation of
$2\Gamma(\rho \rho \to 2\pi^+ 2\pi^-)=\Gamma(\rho \rho \to \pi^+ \pi^- 2\pi^0 )$ as listed in Table~\ref{Tab:4pi}.

Hence, if the structure of $f_0 (1370)$ is more likely to decay via the $\sigma \sigma$ intermediate state, 
the width of the $2 \pi^+ 2 \pi^-$ channel will be approximately twice that of the $\pi^+ \pi^- 2 \pi^0$ channel. 
In contrast, if the $\rho \rho$ intermediate state dominates, the width of the $\pi^+ \pi^- 2 \pi^0$ will be almost twice that of the $2 \pi^+ 2 \pi^-$ channel.
Including both types of processes, the ratio of $\Gamma_{2 \pi^+ 2\pi^-} / \Gamma_{ \pi^+ \pi^- 2\pi^0}$ falls between these two limits.
Comparing this ratio to experimental measurements can help clarify the decay mechanism of $f_0(1370)$, examining our scenario. 
Similarly, the ratio between the widths of the $4 \pi^0$ state and the other $4 \pi$ states can be used to estimate the contributions from the $\rho \rho$ intermediate state, 
as only the $\sigma \sigma$ intermediate state can contribute to the $4 \pi^0$ final state.

\begin{figure}
        \centering
       \subfigure[]{\includegraphics[width=2.8 in]{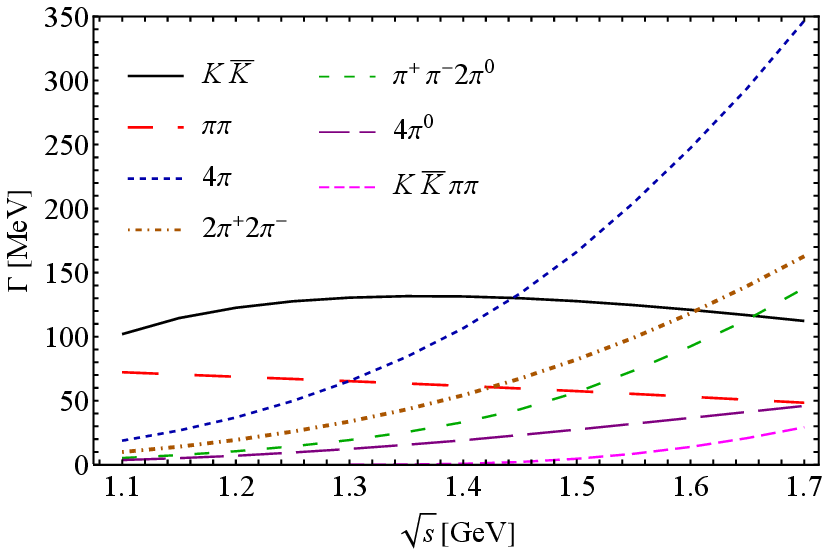}} \qquad \qquad 
       \subfigure[]{\includegraphics[width=2.8 in]{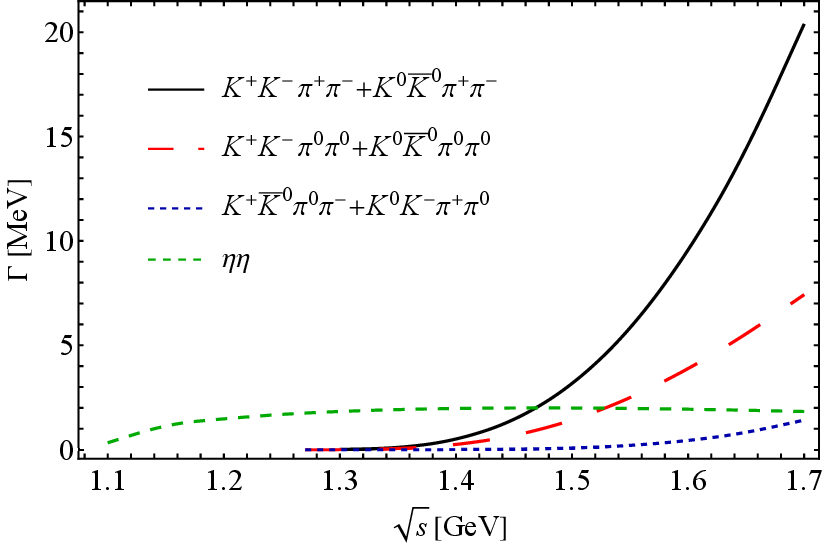}}  
      \caption{The dependence of the partial widths of different channels on the center-of-mass energy $\sqrt{s}$
      with fixed coupling $g_{F\kappa\bar{\kappa}}=35$ GeV and cut-off parameter $\alpha(\alpha_B)=2(1)$.
       Because the scales in widths for some channels are much smaller than others, we present them in two separate figures with distinct scale for clarity.
      (a) $K\bar{K}$, $\pi \pi$, $4 \pi$ and $K \bar{K} \pi \pi$ channels. Different specific $4 \pi$ states are also presented.
       (b) $K \bar{K} \pi \pi $ and $\eta \eta $ channels. Different specific $K \bar{K} \pi \pi$ states are also presented. Note that $\Gamma(K^+ K^- \pi^+ \pi^-)= \Gamma(K^0 \bar{K}^0 \pi^+ \pi^-)$,
 $\Gamma(K^+ \bar{K}^0 \pi^0 \pi^-)=\Gamma(K^0 K^- \pi^+ \pi^0)$ and 
 $\Gamma(K^+K^- \pi^0 \pi^0 )=\Gamma(K^0 \bar{K}^0 \pi^0 \pi^0)$. For the $\eta \eta$ channel, $\Gamma(F\to \eta \eta )$ is presented.}  \label{Fig:DependentOnMs}
\end{figure}

The decay pattern for a broad state is not as accurate to define at a fixed energy as the narrow state, owing to its wide span mass distribution.
The partial widths of some channels might be sensitive to the center-of-mass energy $\sqrt{s}$ over its distribution where some relevant thresholds are located in. 
Hence, the well-defined branching ratios of different decay channels for a broad state, are generally obtained by integrating over its mass distribution, 
a process typically done in the experimental analysis.
For a further reliable study on the decay pattern of the $f_0(1370)$, we present
the dependence of the partial widths of different channels on the center-of-mass energy $\sqrt{s}$.
We plot the partial widths of different channels varying along with the energy $\sqrt{s}$
with fixed coupling $g_{F\kappa\bar{\kappa}}=35$ GeV and cut-off parameter $\alpha(\alpha_B)=2(1)$ in Fig.~\ref{Fig:DependentOnMs}.

\begin{figure}
        \centering
       \subfigure[]{\includegraphics[width=2.8 in]{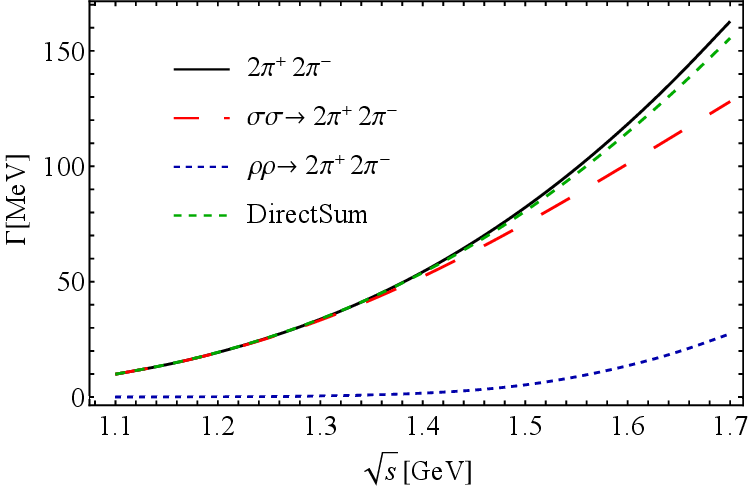}} \qquad \qquad 
       \subfigure[]{\includegraphics[width=2.8 in]{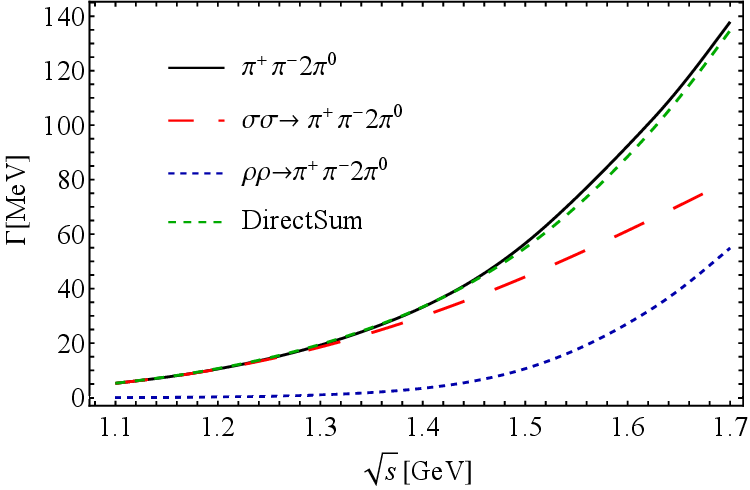}}  
      \caption{ The dependence of the partial widths of $4 \pi $ states on the center-of-mass energy $\sqrt{s}$.
       Contributions from different intermediate channels $\sigma \sigma$ and $\rho \rho$ are also plotted.  
        (a) For the $2 \pi^+ 2\pi^-$ states. (b)  For the $\pi^+ \pi^- 2 \pi^0$ states.   }  \label{Fig:Distinctrhosigma}
\end{figure}

As we can see from Fig.~\ref{Fig:DependentOnMs}(a) and (b), with the increase of $\sqrt{s}$ from $1.1$ GeV to $1.7$ GeV,
the decay widths of the two-body channels such as $K \bar{K}$, $\pi \pi$ and $\eta \eta  $ are stable.
 While for the four-body decay channels $4 \pi$ and $K \bar{K} \pi \pi$,  the decay width continuously increase along with the increase of $\sqrt{s}$.
 The dramatic increase in width of the $K \bar{K} \pi \pi$ channel 
 is more evident in Fig.~\ref{Fig:DependentOnMs}(b) than in Fig.~\ref{Fig:DependentOnMs}(a) due to the scales.
 This phenomenon can be attributed to the increased phase space for the four-body decay and the threshold open for the $\kappa \bar{\kappa}$ and $\rho \rho$ intermediate states as $\sqrt{s}$ increases.
 The $\rho \rho$ threshold open effect is apparent as shown in Fig.~\ref{Fig:Distinctrhosigma}, which will be discussed later.
As a consequence of these threshold effects, the apparent lineshape of the resonance varies depending on the specific decay channel, consistent with the analysis presented in~\cite{Cheng:2024sus}.
 From another perspective, the analytic continuation of a system with pole deep inside the complex plane can be highly non-trivial,
 and the projection onto the real axis, where experiments are performed, can differ widely depending on the complexity of the system.
In Fig.~\ref{Fig:DependentOnMs} we can see that before energy point about $1.45$ GeV, the largest contribution to the total width is from the $K \bar{K}$ channel, 
the order of these partial widths is $\Gamma_{K \bar{K}} > \Gamma_{4 \pi} > \Gamma_{\pi \pi} > \Gamma_{\eta \eta}>\Gamma_{K \bar{K} \pi \pi}$.
 When $\sqrt{s}$ is larger than $1.45$ GeV, the partial width of the $4 \pi$ channel exceeds that of the $K \bar{K}$ channel, becoming the most dominant contribution. However, when $\sqrt{s}$ is less than about $1.34$ GeV, the partial width of $4 \pi $ channel is even smaller than that of the $\pi \pi$ channel. 
Similarly, along with the increase of $\sqrt{s}$, 
the large difference between the partial widths of the $K \bar{K}\pi \pi$ and $\pi \pi$ channels at small $\sqrt{s}$ start to reduce gradually.


Analysis on the $2\pi^+ 2\pi^-$ spectrum in~\cite{Gaspero:1992gu} gives the ratio of $Br(\rho^0 \rho^0 \to 2\pi^+ 2\pi^- ) : Br(\sigma \sigma \to 2\pi^+ 2\pi^- )$ as $3:7$.
Our corresponding result shown in Table~\ref{Tab:4pi} yields a much smaller $\rho\rho$ fraction, with $\rho^0\rho^0:\sigma\sigma \approx 1:38$ and $\approx 1:12$ for the cut-off choices $\alpha (\alpha_B)=2(1)$ and $3(2)$, respectively.
Our calculations indicate that the strong suppression of the $\rho\rho$ channel relative to $\sigma\sigma$ is a distinct feature of the $\kappa\bar{\kappa}$ molecular picture.
We note that our result is obtained at fixed energy $\sqrt{s}=1.37$ GeV, while the experimental branching ratios are obtained by integrating over the mass distribution of $f_0(1370)$.
The ratio is also sensitive to the parametrization of the $\sigma$, which we adopted differently from~\cite{Gaspero:1992gu}.

For a more reliable comparison, we plot the respective partial widths of these two parts varying 
along with the center-of-mass energy $\sqrt{s}$ in Fig.~\ref{Fig:Distinctrhosigma}, to investigate the resulting trend of their ratio after performing the mass distribution integration.
Fig.~\ref{Fig:Distinctrhosigma}(a) and (b) refer to the $2 \pi^+ 2 \pi^-$ and $\pi^+ \pi^- 2 \pi^0$ states, respectively.
The black solid line indicates the total width of the specific $4 \pi$ state, which includes the interference between the $\rho \rho$ and $\sigma \sigma$ intermediate states.   
The red long dashed line and the blue short dashed line indicate the exclusive widths from the $\sigma \sigma$ and $\rho \rho$ intermediate states respectively.
In Fig.~\ref{Fig:Distinctrhosigma}(a), we observe that the increase in the $\sigma \sigma \to 4\pi$ process is moderate,
whereas the $\rho \rho \to 4 \pi$ process experiences a rapid increase once the energy $\sqrt{s}$ surpass the $\rho \rho $ threshold.
Nevertheless, even after considering the integration over the whole mass distribution, the $\rho\rho$ contribution remains insufficient compared to the ratios from experiments.
The $\rho\rho$ contributions observed in experiments in $f_0(1370)$ decay to $4 \pi$ may largely originate from the nearby $f_0(1500)$,
which is suggested to be a $\rho\rho$ molecular state~\cite{Wang:2019niy} and is expected to decay dominantly via the $\rho\rho$ channel.
If the experimental analyses, typically relying on sums of Breit-Wigner functions, do not adequately separate these two broad overlapping resonances,
the $\rho\rho$ events from $f_0(1500)$ decay would be misattributed to $f_0(1370)$, leading to an overestimation of the $\rho\rho$ branching fraction of $f_0(1370)$. 
  
We also plot the direct sum of the exclusive widths of the $\rho \rho$ and $\sigma \sigma $ cases, which is indicated by green dashed line and marked as ``DirectSum'' in Fig.~\ref{Fig:Distinctrhosigma}.
We find that it is almost coincides with the black solid line along all the $\sqrt{s}$ for both the $2 \pi^+ 2\pi^-$ and $\pi^+ \pi^- 2\pi^0$ channels,
which implies that the $\rho \rho$ and $\sigma \sigma$ channels are almost orthogonal and their interference is negligible. 
This is because the $\rho \pi \pi$ vertex is of $P$-wave while the $\sigma \pi \pi$ vertex is of $S$-wave.
 Such a phenomenon is also confirmed in the analysis in Ref.~\cite{Anisovich:2001jb}.

\begin{table}
        \caption{Collection of the available experiments and our results in the case of $\alpha(\alpha_B)=1(0.5)/\mathbf{2(1)}/3(2)$.
        The coupling $g_{F\kappa \bar{\kappa}}$ is fixed to be $35$ GeV. }\label{Tab:channelratios}
        \begin{tabular}{l | c |  c  }
                \hline  \hline
                 Expression                                                                    & \quad  Exp. \quad                                          &  \quad Theo. (Fixed $\sqrt{s}=1.37$ GeV ) \quad \\
                \hline 
                       $\Gamma(F \to K\bar{K}) / \Gamma_F$                                     &        $0.35 \pm 0.13$\cite{Bugg:1996ki}                                                                                                              &       $0.35/\mathbf{0.45}/0.49$                                                   \\
                       $\Gamma(F \to \pi \pi ) / \Gamma_F  $                                   &        $0.26 \pm 0.09$ \cite{Bugg:1996ki};~ $< 0.10$ \cite{Ochs:2013gi}                                                                                                          &       $ 0.16 /\mathbf{0.22}/0.25 $                                                \\
                       $\Gamma(F \to 4 \pi ) /\Gamma_F$                                        &         $>0.72$ \cite{Gaspero:1992gu}                                                               &       $ 0.46/\mathbf{0.32}/0.25 $                                                  \\
                       $\Gamma(F \to K \bar{K} \pi \pi) / \Gamma_{F}$                          &                 \text{No data}                                                                                                                        &       $2.5 \times 10^{-3}/\mathbf{1.1\times 10^{-3} }/ 7.9 \times 10^{-4}$         \\
                       $\Gamma(F \to K \bar{K}) / \Gamma(F\to \pi \pi )$                       &     $0.08 \pm 0.08 $\cite{BES:2004twe};~$0.46 \pm 0.19$\cite{WA102:1999fqy};~  $0.91 \pm 0.2 $ \cite{OBELIX:2002lhi};~                                &       $2.15/\mathbf{2.10}/2.01$                                                    \\
                      $\Gamma(F\to \eta \eta)/\Gamma(F \to 4 \pi)   $                          & $ (4.7 \pm 2.0) \times 10^{-3}$\cite{WA102:2000lao};~$(28 \pm 11)\times 10^{-3}$\cite{Anisovich:2001ay}                                               &       $ 11.7 \times 10^{-3} / \mathbf{ 21.0 \times 10^{-3}} / 29.3 \times 10^{-3}  $     \\
                    $\Gamma'(F\to \eta \eta)/\Gamma(F \to 4 \pi)$     & $ (4.7 \pm 2.0) \times 10^{-3}$\cite{WA102:2000lao};~$(28 \pm 11)\times 10^{-3}$\cite{Anisovich:2001ay}                                               &     $ 5.4\times 10^{-4} / \mathbf{ 9.7 \times 10^{-4} } / 1.3 \times 10^{-3} $         \\
                      
                     $\Gamma(F \to 4 \pi ) / \Gamma(F\to \pi \pi )$                            & $      0.3 \pm 0.12$\cite{Albaladejo:2008qa};~ $0.1\sim0.25$\cite{Bugg:2007ja}                                                                        &       $  2.79 /\mathbf{1.48} /1.02$         \\
                      $\Gamma(F \to \rho \rho ) / \Gamma(F \to 4 \pi)$                         & $0.26\pm 0.07$\cite{CRYSTALBARREL:2001ldq}                                                                                                            &       $ 0.02 /\mathbf{0.04}/0.11 $   \\

$\Gamma(F \to \sigma \sigma ) / \Gamma (F \to 4 \pi)$  &  $0.51 \pm 0.09$ \cite{CRYSTALBARREL:2001ldq}         &  $0.79 / \mathbf{0.78}/ 0.74$ \\
 $\Gamma(F \to 2 \pi^+ 2 \pi^- )/ \Gamma(F \to 4 \pi  )$        & $0.42 \pm 0.014$\cite{Gaspero:1992gu}   &       $  0.51 / \mathbf{0.51} /0.50$  \\
  $\Gamma(F \to  \pi^+ \pi^- 2 \pi^0 )/ \Gamma(F \to 4 \pi  )$       &  $0.512 \pm 0.019$\cite{Gaspero:1992gu}   &  $ 0.31 /\mathbf{0.31} /0.33$   \\
 $\Gamma(F \to \sigma \sigma \to 4 \pi^0 )/ \Gamma(F \to 4 \pi  )$     & $0.068 \pm 0.005$ \cite{Gaspero:1992gu};~  $0.27 \pm 0.06$\cite{CRYSTALBARREL:2001ldq}                                                                                                    &       $ 0.18/\mathbf{0.18}/0.18$   \\
                  \hline \hline
  \end{tabular}
\end{table}

In Table~\ref{Tab:channelratios}, we collect the current available data and our theoretical results of 
the ratios between different decay channels as well as the ratios of the partial widths to the total width, 
which are calculated at fixed energy $\sqrt{s}=1.37$ GeV. 
Again one should be cautious with the fact that some ratios are obtained by integrating over the mass distribution of $f_0(1370)$.

 From Table~\ref{Tab:channelratios} we can see that for those decay channels ``seen'' in $f_0(1370)$ decay, no single experimental number on branching ratios nor ratios thereof has been determined because of conflicting results. 
This is quite different from the nearby $f_0(1500)$ with five well established branching ratios.
As stated in~\cite{Pelaez:2022qby}, generally $f_0(1370)$ analyses suffer from some aspects of the ``model problem'':
parameterization choices, most frequently BW, K-matrices, non-resonant backgrounds and isobars,
a priori selection of decay channels, two body or quasi-two-body approximations, etc.
Hence, most of the current data are model-dependent.
Indeed, shape of such resonances changes with the dynamics of the process where they appear,
therefore they must be rigorously identified from the process-independent associated poles.

In Table~\ref{Tab:channelratios}, the relation between the $K\bar{K}$ and $\pi \pi$ channels has  two kinds of results. Qualitatively, measurements from the $\pi \pi$ and $\pi p$ scattering obtained at the mass of $f_0(1370)$~\cite{Bugg:1996ki} support that $\Gamma(F \to K \bar{K})$ is larger than $\Gamma(F \to \pi \pi)$, which is consistent with
our results at $\sqrt{s}=1.37$ GeV.
In contrast, measurements from the $J/ \psi$ hadronic decay~\cite{BES:2004twe}, $J/\psi$ radiative decay~\cite{Sarantsev:2021ein}, the $pp$ central production~\cite{WA102:1999fqy} and the $p \bar{p}$ annihilation at rest~\cite{OBELIX:2002lhi} show an inverse branching fraction ratio.
As mentioned in the Introduction, the data of $J /\psi$ hadronic decay recoiling against $\phi$ even gives a very small value for this ratio of $0.08 \pm 0.08$ due to the conspicuous signal in $\pi \pi$ but absence of any corresponding peak in $K^+ K^-$.
We attempt to explain this by the destructive interference between the two parts in the $f_0(1370)$ wavefunction: $s\bar{s}$ and the molecular component $\kappa \bar{\kappa}$. 
Although the $s \bar{s}$ component is not included in the wavefunction of Eq.~(\ref{Eq:wavefunction}) in our calculation above, a pure molecular state is unlikely to exist in reality, and the $f_0(1370)$ wavefunction could include a small $s\bar{s}$ component.
$s\bar{s}$ is produced in $J/\psi$ decay recoiling against $\phi$ as leading order and it strongly couple to $K \bar{K}$ but contribute to $\pi \pi$ with significantly weaker strength, leading to a possible destructive effect in the $K\bar{K}$ channel but not affecting the $\pi \pi$ channel. Such an $s\bar{s}$ destructive effect might also appear in the $ p p $ central production and $J/\psi$ radiative decay but appear in the $\pi p $ scattering with a smaller fraction.
The more quantitative discussion about this issue is complex and needs more reliable analyses on experiments.

 With the coupling $g_{\kappa K \eta},
$ the theoretical ratio of the widths between $\eta \eta$ and $4 \pi$ channels is consistent with the data in Ref.\cite{Anisovich:2001ay}, while with the coupling $g'_{\kappa K \eta}$, this ratio is more consistent with the data in Ref.~\cite{WA102:2000lao}.
Whether the $4 \pi$ channel is the most dominant one is also conflicting from the experimental ratios in Table~\ref{Tab:channelratios}.
The accurate analyses on the $4 \pi$ spectra are difficult because it is hard to disentangle the contribution from $f_0(1370)$ and $f_0(1500)$
due to the subtle parametrization of the broad $f_0(1370)$. 
In most earlier work, the $f_0(1370)$ has been fitted with a Breit-Wigner amplitude of constant width.
This is inaccurate because the resonance couples to $4 \pi$ channel, which give rise to a strongly $s$-dependent width due to the rapid opening 
of the $4 \pi$ thresholds for $\sigma \sigma $ and $\rho \rho$. As Achasov and Shestakov have emphasized~\cite{Achasov:1995nh},
if one describes these spectra by an $f_0$ state with an energy-dependent total width its mass has to lie above $1500$ Mev,
and even no appreciable coupling of $f_0(1370)$ to the $4 \pi$ channel is required. 
While we should still keep skeptical about this conclusion because it is also drawn based on the Breit-Wigner like parametrization.

The analysis in Ref.~\cite{Chen:2000dra} carefully considered the Bose symmetry interference effects 
   and found that these effects cause the ratios between different specific $4 \pi$ states to obviously differ
   from the naive counting values without considering these effects, as what is evaluated in~\cite{Gaspero:1992gu} based on the $Br(2 \pi^+ 2\pi^-)$ to obtain these ratios.
   Hence, the estimated ratios of $ Br(2 \pi^+ 2\pi^-): Br(\pi^+ \pi^- 2 \pi^0): Br(4 \pi^0)$ in~\cite{Gaspero:1992gu} is not reliable.
While such effects have been fully considered in our calculations.

\section{Summary} \label{Sec:summary}

In summary, motivated by the fact that the $f_0(980)$ and $f_0(1790)$ are respectively plausible $K \bar{K}$ and
$K^* \bar{K}^*$ molecules  and the phenomenology that the $f_0(980)$, $f_0(1370)$ and $f_0(1790)$ appear successively in the $\pi \pi$ spectrum in $J/\psi$ hadronic decay recoiling against $\phi$, where the $s\bar{s}$ is produced in leading order, we assume that $f_0(1370)$ is a $\kappa \bar{\kappa}$ molecular state and 
 calculate the partial widths of various channels in $f_0(1370)$ decay, including $K \bar{K}$, $\pi \pi$, $\eta \eta$, $4 \pi$ and $K \bar{K} \pi \pi$.
We treat the coupling $g_{F \kappa \bar{\kappa}}$ ($F$ indicates $f_0(1370)$) as a free parameter and fit it to the measured total width of $f_0(1370)$ listed in the RPP,
 finding that this coupling should be in the range of $25 \sim 40$ GeV with cut-off parameter $\alpha(\alpha_B) =2(1)\sim 3(2)$.
 The decay pattern is independent of this coupling because it is a common coupling for all the decay channels.

 At the nominal mass of $f_0(1370)$: $m_F=1.37$ GeV and with $g_{F \kappa \bar{\kappa}}=35$ GeV and $\alpha(\alpha_B)=2(1)$, the $K \bar{K}$ channel has the largest partial width, followed by the $4\pi$ and $\pi \pi$ channels. The $\pi \pi$ and $4 \pi$ channels have comparable partial widths. These three channels are the most dominant decay channels in $f_0(1370)$ decay.
 Different strategies for determining $\kappa K \eta$ coupling result in significant variations in the partial width of the $\eta \eta$ channel. 
 We have found the distinct decay properties in $\eta \eta $ channel between $\kappa \bar{\kappa}$ and $K^* \bar{K^*}$ molecules, which can be examined in the BESIII data of $J/\psi \to \phi \eta \eta$. 
The partial width of the $K\bar{K} \pi \pi$ channel is about $300$ keV in total.
 Notice that although the $K\bar{K} \pi \pi$ channel proceeds through tree-level decay, while it has very limited four-body phase space around $1.37$ GeV.

We also test the dependence of the partial widths of different channels on the center-of-mass energy $\sqrt{s}$. 
The decay widths of the two-body channels such as $K \bar{K}$, $\pi \pi$ and $\eta \eta$ are stable with the variation of $\sqrt{s}$. 
While the decay widths of the four-body decay channels $4 \pi$ and $K \bar{K} \pi \pi$ continuously increase along with the increase of $\sqrt{s}$.
With $\sqrt{s}$ larger than $1.45$ GeV, the partial width of the $4 \pi$ channel exceed that of the $K\bar{K}$ channel, becoming the most dominant decay channel. Such phenomenon might cause different manifestation of $f_0(1370)$ in different spectra.  

Not all of the ratios reflecting the decay pattern in our calculations are consistent with the current data,
such as the $4 \pi$ dominant conclusion and the ratio of $K\bar{K}$ over $\pi \pi$ in some measurements.
However, in $f_0(1370)$ decay, most of the current data are model-dependent, relying on fitting sums of Breit-Wigner functions 
and no single experimental number on branching ratios nor ratios thereof are determined because they are conflicting. 
Hence, these inconsistencies can not rule out the $\kappa \bar{\kappa}$ assignment for $f_0(1370)$.
The small $K \bar{K}$ branching ratio due to the absence of corresponding peak in $K\bar{K}$ spectrum,
such as in $J/ \psi$ hadronic decay recoiling against $\phi$, contradict with our result. We attempt to explain that by the destructive interference in decay to $K\bar{K}$ between the $s\bar{s}$ and $\kappa \bar{\kappa}$ components in the $f_0(1370)$ wavefunction. 
The $4 \pi$ spectra is also elusive because it is hard to disentangle the contribution from $f_0(1370)$ and $f_0(1500)$ due to the strongly $s$-dependent
width of $f_0(1370)$.
The ratios between different $4 \pi$ states can be related to decay mechanisms as the intermediate states $\rho \rho$ and $\sigma \sigma$ have distinct features in $4 \pi$ decay.
In our $\kappa\bar{\kappa}$ molecular picture, the $\sigma\sigma$ channel dominates over $\rho\rho$, which is discrepant to current data.
The $\rho\rho$ signal observed in the $f_0(1370)$ region may largely originate from the nearby $f_0(1500)$, 
a $\rho\rho$ molecular candidate whose $\rho\rho\to4\pi$ events can be misattributed to $f_0(1370)$ in experimental analyses that do not adequately separate these two broad overlapping resonances.

This work relies on a number of model assumptions that should be kept in mind when interpreting the results:
 (i) the pure $\kappa\bar{\kappa}$ molecular ansatz of Eq.~(\ref{Eq:wavefunction}) without $s\bar{s}$ or other mixing components.
  (ii) the simple-pole propagators $1/(s-s_{\text{pole}})$ for the broad intermediate states $\kappa$ and $\sigma$.
 (iii) the approximate SU(3) flavor symmetry relations applied to the lowest scalars.  
 (iv) the reliance on particular phenomenological determinations of couplings from the literature, which carry their own model dependence and uncertainties. 
 (v) the coupling estimation methods such as the spectator approximation used to estimate the couplings involving $\kappa$ and $\sigma$. 
 (vi) the specific choice of the monopole form factor and the cut-off parameter for the exchanged states.
 Despite these limitations, the model provides useful qualitative insights into the decay pattern expected for a $\kappa\bar{\kappa}$ molecular interpretation of $f_0(1370)$.

We need further reliable analyses on the $f_0(1370)$ in both theoretical and experimental aspects.
 The BESIII Collaboration has recorded data with significantly improved quality and statistics. 
They can attempt to search for the $f_0(1370)$ signal in the $K \bar{K} \pi \pi / 4 \pi /\eta \eta $ channels in $J/\psi$ hadronic decay recoiling $\phi$. 
 Such measurements can help us to test the 
$\kappa \bar{\kappa}$ molecule assignment for $f_0(1370)$.
 The pattern for the different specific $K \bar{K} \pi \pi / 4 \pi$ states also has unique feature in this scenario, which can be used as a further check.

\section*{ACKNOWLEDGMENTS}

The authors appreciate the useful discussions with Feng-Kun Guo, Jia-Jun Wu, Xiong-Hui Cao, Shu-Ming Wu, Zhen-Hua Zhang, Zhao-Sai Jia, and Xu Zhang.

\appendix 

\section{Isospin conventions}\label{Appendix:A}
Let $I$ being an isospin operator in the Hilbert space, and its corresponding isospin transform matrix is $\bm{\tau}/2$.  
As discussed in Ref.~\cite{Martin:1970hmp}, due to the fact that $\hat{C}$ does not commute with the step operators $I_1 \pm i I_2$,
the relative phases of the antiparticle states will not satisfy the following requirement
\begin{eqnarray}
        (I_1 \pm i I_2) | I m \rangle = + \sqrt{(I \mp m)(I \pm m+1)}|I, m\pm 1 \rangle.
\end{eqnarray}
Both of the deficiencies are removed by constructing the antiparticle isospin multiplet, which is denoted by $\overline{| I m \rangle}$ ,
 with the help of the G-conjugation operator:
\begin{eqnarray}\label{Eq:isospin-anti}
        \overline{| I m \rangle} &\equiv& \eta_I \hat{G} |I m\rangle, \\ \nonumber
                                 &=& \eta_I (-1)^{I-m} \hat{C}|I,-m \rangle. 
\end{eqnarray}
$\eta_I$ is choose to be $-1$ as discussed in~\cite{Martin:1970hmp} .

We take the convention as $\kappa^+= | 1/2, 1/2 \rangle$, $\kappa^0=|1/2, -1/2 \rangle$, 
and we construct the isospin states of antiparticles using the prescription given by Eq.~(\ref{Eq:isospin-anti}).
It is a natural choice to take positive charge parity for $\kappa$ particles as the neutral scalars $0^{++}$:  
\begin{eqnarray}
        \hat{C} |\kappa^+ \rangle= + |\kappa^-\rangle, \qquad \hat{C}|\kappa^0\rangle =+ |\bar{\kappa}^0 \rangle
\end{eqnarray}

\begin{eqnarray}
        |1/2, -1/2 \rangle &=&(-1) \times (-1)^{1/2 - (-1/2)} \hat{C} |1/2, 1/2 \rangle    \\
                           &=& \hat{C}|\kappa^+\rangle \nonumber \\
                           &=& + |\kappa^- \nonumber \rangle
\end{eqnarray}

\begin{eqnarray}
        |1/2, 1/2 \rangle &=&(-1) \times(-1)^{1/2 - (1/2)} \hat{C} |1/2, -1/2 \rangle  \\
                           &=& -\hat{C}|\kappa^0\rangle \\
                           &=& -|\bar{\kappa}^0\rangle
\end{eqnarray}
Hence, the antiparticle isospin states are $\bar{\kappa}^0=-|1/2, 1/2 \rangle$,  $\kappa^-=|1/2,-1/2 \rangle$.
 The adopted convention in our work of the isospin states of the particles is shown in Table~\ref{Tab:isospin}.

\begin{table}
        \centering
        \caption{Conventions for isospin states and flavor wave functions of the particles involved in this work.}
        \begin{tabular}{c c c}
                \hline \hline
                Particles  &  Isospin state  &  Flavor wave-function    \\
                \hline
                  $\pi^+$ & $ -| 1, 1 \rangle$  &  $u \bar{d}$ \\
                  $\pi^0$ & $|1, 0 \rangle $ &$( 1/\sqrt{2})(u \bar{u}-d \bar{d})$  \\
                  $\pi^-$ &  $| 1,-1 \rangle$ & $ d \bar{u} $ \\
                  $\kappa^+  /  K^+$ & $|1/2, 1/2 \rangle$ & $u \bar{s}$ \\
                  $\kappa^0 / K^0  $ & $|1/2, -1/2 \rangle$ & $d \bar{s}$ \\
                  $\bar{\kappa}^0 / \bar{K}^0$ & $-|1/2, 1/2 \rangle$ & $s \bar{d}$ \\
                  $\kappa^- /K^- $  & $|1/2, -1/2 \rangle$ & $s \bar{u}$ \\
                  \hline \hline 
        \end{tabular}\label{Tab:isospin}
\end{table}

\section{ Estimations of the effective coupling constants.}\label{Appendix:B}

\begin{itemize}
        \item  $\rho \kappa \bar{\kappa}$  
\end{itemize}

 We assume $\kappa$ to be a molecular state consisting of $K \pi$, its couplings to other state can be described as follows:
the $\kappa \to V \kappa / S \kappa$ interaction can be viewed as one of the pseudoscalar components inside $\kappa$ interacting with the vector or scalar, 
while the other component acts approximately as a spectator.
For example, 
the $\kappa \kappa \rho$ vertex can be approximately related to the $\rho \pi \pi$ and $\rho K \bar{K}$ couplings 
by guaranteeing that the $K \pi$ is projected to the isospin state of $\kappa$. 
In the molecular picture, the flavor wave function of $\kappa$ can be expressed as 
\begin{eqnarray}
        \kappa^+ & =& \frac{1}{\sqrt{3}} | K^+ \pi^0 \rangle + \frac{\sqrt{2}}{\sqrt{3}} | K^0 \pi^+ \rangle, \\
         \kappa^0 &=& \frac{\sqrt{2}}{\sqrt{3}} |K^+ \pi^- \rangle - \frac{1}{\sqrt{3}}|K^0 \pi^0 \rangle. \label{Eq:ff-kappa}
\end{eqnarray}
Based on the $g_{K^{*0} K^+ \pi^-} $ we can obtain the $g_{\rho K \bar{K}}$
and $g_{\omega K \bar{K}}$ couplings by the SU(3) flavor symmetry.
By the Lagrangian with SU(3) symmetry, the vertex coupling constants are connected as following~\cite{Cheng:2021nal}:
\begin{eqnarray}
       g_{VPP} &=& g_{K^{*0} K^+ \pi^-} \\
       g_{\rho K^+ K^- }  &=& -g_{VPP}/\sqrt{2},  \\    
        g_{\rho^0 \pi^+ \pi^-} &=& -\sqrt{2} g_{VPP} .
           \label{Eq:su3}
\end{eqnarray} 
The possible interactions between $\rho$ and the pseudoscalar components inside $\kappa^+$ will include $\rho^0 K^+ K^-$, $\rho^0 K^0 \bar{K}^0$ and $\rho^0 \pi^+ \pi^-$. 
The coupling $\rho^0 \kappa^+ \kappa^-$ can be calculated as 
\begin{eqnarray}
        g_{\rho^0 \kappa^+ \kappa^-}&=& \frac{1}{3} g_{\rho^0 K^+ K^-} + \frac{2}{3} (g_{\rho^0 K^0 \bar{K}^0} + g_{\rho^0 \pi^+ \pi^-})    \\
                                          &=& \frac{1}{3} (\frac{-g_{VPP}}{\sqrt{2}}) +  \frac{2}{3} (\frac{g_{VPP}}{\sqrt{2}}-\sqrt{2} g_{VPP})  \nonumber  \\
                                          &=& \frac{-g_{VPP}}{\sqrt{2}} \nonumber \\
                                          &=& g_{\rho^0 K^+ K^-}. \nonumber
\end{eqnarray}
The decomposition of the vertex $\rho^+ \kappa^+ \bar{\kappa}^0$ is
\begin{eqnarray}
       g_{\rho^+ \kappa^+ \bar{\kappa}^0}&=& \frac{\sqrt{2}}{\sqrt{3}} \times \frac{-1}{\sqrt{3}} g_{\rho^+ \pi^+ \pi^0} 
       + \frac{1}{\sqrt{3}} \times \frac{-1}{\sqrt{3}} g_{\rho^+ K^+ \bar{K}^0} + \frac{1}{\sqrt{3}} \times \frac{\sqrt{2}}{\sqrt{3}} g_{\rho^+ \pi^0 \pi^+},
\end{eqnarray}
with SU(3) symmetry, $g_{\rho^+ K^+ \bar{K}^0}= - g_{VPP}$, $g_{\rho^+ \pi^+ \pi^0}= \sqrt{2} g_{VPP} $ and $g_{\rho^+ \pi^0 \pi^+}= -\sqrt{2} g_{VPP}$, 
we have 
\begin{eqnarray}
        g_{\rho^+ \kappa^+ \bar{\kappa}^0} = - g_{VPP}= g_{\rho^+ K^+ \bar{K}^0}.
\end{eqnarray}
Similarly, we have
\begin{eqnarray}
        g_{\rho^0 \kappa^0 \bar{\kappa}^0}&=& \frac{1}{3} g_{\rho^0 K^0 \bar{K}^0} + \frac{2}{3} (g_{\rho^0 K^+ K^-} + g_{\rho^0 \pi^- \pi^+})    \\
                                          &=& \frac{1}{3} \biggl(  \frac{g_{VPP}}{\sqrt{2}} \biggr) +  \frac{2}{3} \biggl(  \frac{-g_{VPP}}{\sqrt{2}} + \sqrt{2} g_{VPP} \biggr)  \nonumber  \\
                                          &=& \frac{g_{VPP}}{\sqrt{2}} \nonumber \\
                                          &=& g_{\rho^0 K^0 \bar{K}^0}. \nonumber
\end{eqnarray}

\begin{itemize}
\item  $\sigma \kappa \bar{\kappa}$
\end{itemize}
    
Regarding the $\sigma$ as a $\pi \pi$ molecular state, its wave function can be written as 
\begin{eqnarray}
        \sigma= \frac{-1}{\sqrt{3}}| \pi^0 \pi^0 \rangle+ \frac{-1}{\sqrt{3}} |\pi^+ \pi^- \rangle +\frac{-1}{\sqrt{3}} |\pi^- \pi^+ \rangle. \label{Eq:ff-sigma}
\end{eqnarray}
A similar decomposition for $g_{\sigma \kappa^+ \kappa^-}$ coupling can be written as:
\begin{eqnarray}
            g_{\sigma \kappa^+ \kappa^-} &=& \frac{1}{3} (2 g_{\sigma \pi^0 \pi^0}) + \frac{1}{3} g_{\sigma K^+ K^-} + \frac{2}{3} g_{\sigma \pi^+ \pi^-} + \frac{2}{3} g_{\sigma K^0 \bar{K}^0}  \nonumber \\
                                         &= & 3 g_{\sigma K^+ K^-}.
\end{eqnarray}

\begin{eqnarray}
         g_{\sigma \kappa^0 \bar{\kappa}^0} &=&\frac{1}{3} (2 g_{\sigma \pi^0 \pi^0}) + \frac{1}{3} g_{\sigma K^0 \bar{K}^0}+\frac{2}{3} g_{\sigma \pi^+ \pi^-}+ \frac{2}{3} g_{\sigma K^+ K^-}. \nonumber \\
                                   &=& 3 g_{\sigma K^0 \bar{K}^0}.
\end{eqnarray}
The factor 2 multiplied to the $g_{\sigma \pi^0 \pi^0}$ account for two different contractions due to the identical particles $\pi^0$.

\begin{itemize}
\item $\sigma \kappa \bar{K}^*$
\end{itemize}

Similarly, one of the components inside the $\sigma$ is assumed to act as a spectator in the interaction,
 then the coupling of $\sigma \kappa \bar{K}^*$ vertex can be approximately related to the coupling of $K^* K \pi$ vertex.
For example, based on the wave functions of $\sigma$ and $\kappa$, the coupling $g_{\sigma \kappa^+ K^{*-}}$ can be expressed as
   
\begin{eqnarray}
        g_{\sigma \kappa^+ K^{*-}} &=& \frac{-1}{\sqrt{3}}\times \frac{1}{\sqrt{3}} \cdot g_{K^{*-} K^- \pi^0}- \frac{1}{\sqrt{3}}\times \frac{\sqrt{2}}{\sqrt{3}} \cdot g_{K^{*-} \bar{K}^0 \pi^-},    \nonumber \\
                                   &=& \frac{-1}{3}\times (\frac{-g_{VPP}}{\sqrt{2}})-\frac{\sqrt{2}}{3}(-g_{VPP}) \nonumber \\
                                   &=& \frac{g_{VPP}}{\sqrt{2}},
\end{eqnarray}
and the $g_{\sigma \kappa^0 \bar{K}^{*0}}$ can be expressed as
\begin{eqnarray}
       g_{\sigma \kappa^0 \bar{K}^{*0}}&=& \frac{-1}{\sqrt{3}}\times \frac{-1}{\sqrt{3}}\cdot g_{\bar{K}^{*0} \bar{K}^0\pi^0} - \frac{1}{\sqrt{3}} \times \frac{\sqrt{2}}{\sqrt{3}} \cdot  g_{\bar{K}^{*0} K^-\pi^+}  \nonumber \\
                                 &=&  \frac{1}{3}\cdot \frac{g_{VPP}}{\sqrt{2}} -   \frac{\sqrt{2}}{3} \cdot (-g_{VPP}) \nonumber \\
                                  &=& \frac{g_{VPP}}{\sqrt{2}}.
                                \end{eqnarray}

\begin{itemize}
\item $\rho \kappa \bar{K}^*$
\end{itemize}
   
 Assuming $g_{\sigma \pi \pi }= g_{\sigma \rho \rho }$ due to the identical flavor wave functions for $\pi$ and $\rho$ and 
based on the flavor SU(3) symmetry, we have  $g_{\sigma \rho^+ \rho^-}= \sqrt{2} g_{VVS} = -0.79+ 2.84i $ GeV  and $g_{\rho^0 \kappa^+ K^{*-}}= g_{VVS}/ \sqrt{2}=-0.39 +1.42 i $.

\section{ Issues about the $\eta$-$\eta'$ mixing  } \label{Appendix:C}

If the $\eta$ state is treated as a pure octet $\eta_8$ in Ref.~\cite{Wang:2021jub}, where the octet matrix 

\begin{equation}
  P_8=\left(
    \begin{array}{ccc}
        \frac{1}{\sqrt{2}} \pi^0 + \frac{1}{\sqrt{6}} \eta_8              & \pi^{+}                      & K^{+}\\
      \pi^{-}         &    -\frac{1}{\sqrt{2}} \pi^0 + \frac{1}{\sqrt{6} } \eta_8  & K^{0} \\
      K^{-} & \bar{K}^{0}&  -\frac{2}{\sqrt{6}} \eta_8 \\
    \end{array}
\right)
\end{equation}
is used. 

However, the SU(3) flavor symmetry is slightly broken, which leads to the mixing of the singlet $\eta_1$ and the octet $\eta _8$. 
 Hence, the physical states $\eta$ and $\eta'$ are the outcomes of the mixing of $\eta_1$ and $\eta_8$.
 Defining a mixing angle $\theta$ in the $\eta_8$-$\eta_1$ basis, the the mixing scheme is 
 \begin{eqnarray}
     \left( \begin{array}{c}
             \eta \\
             \eta'
      \end{array} \right )=  \left( \begin{array}{cc}
             \cos \theta  &  -\sin \theta \\
              \sin \theta & \cos \theta 
      \end{array} \right )  \left( \begin{array}{c}
             \eta_8 \\
             \eta_1
      \end{array} \right ),
 \end{eqnarray}
and
  \begin{eqnarray}
     \left( \begin{array}{c}
             \eta_8 \\
             \eta_1
      \end{array} \right )=  \left( \begin{array}{cc}
             \cos \theta  &  \sin \theta \\
             - \sin \theta & \cos \theta 
      \end{array} \right )  \left( \begin{array}{c}
             \eta \\
             \eta'
      \end{array} \right ). \label{Eq:mixing}
 \end{eqnarray}
 Incorporating the mixing, a U(3) flavor symmetry matrix which is the sum of the octet and singlet matrices can be adopted as shown in~\cite{Mathieu:2010ss}, i.e.,
\begin{eqnarray}
   U= P + \frac{\eta_1}{\sqrt{3}}  \mathds{1}_{3 \times 3 }= \left(
    \begin{array}{ccc}
        \frac{1}{\sqrt{2}} \pi^0 + \frac{1}{\sqrt{6}} \eta_8 + \frac{1}{\sqrt{3}}\eta_1            & \pi^{+}                      & K^{+}\\
      \pi^{-}         &    -\frac{1}{\sqrt{2}} \pi^0 + \frac{1}{\sqrt{6} } \eta_8 + \frac{1}{\sqrt{3}}\eta_1  & K^{0} \\
      K^{-} & \bar{K^{0}}&  -\frac{2}{\sqrt{6}} \eta_8 + \frac{1}{\sqrt{3}}\eta_1\\
    \end{array} 
\right).
    \end{eqnarray}
For determining the coupling of $\eta$, we can replace the $\eta_8$ and $\eta_1$ in terms of $\eta$ and $\eta'$ based on Eq.~(\ref{Eq:mixing}).

Expanding the effective Lagrangian $\mathcal{L}= g_{SPP} \langle SPP \rangle $ with the pseudoscalar matrix $U$ under the U(3) flavor symmetry, we have 
\begin{eqnarray}
    \mathcal{L}_{\kappa K \eta }&=& g_{SPP} \bigg[ (\frac{1}{\sqrt{6} }- \frac{2}{\sqrt{6}}) \kappa K \eta_8 +  \frac{2}{\sqrt{3}} \kappa K \eta_1 \bigg] \nonumber \\
      &=& g_{SPP} \bigg[\frac{-1}{\sqrt{6}} \cos\theta + \frac{2} {\sqrt{3}}(-\sin \theta) \bigg]\kappa K \eta 
\end{eqnarray}
and 
\begin{eqnarray}
      \mathcal{L}_{\kappa K \pi^+ }= g_{SPP} \kappa K \pi^+.
\end{eqnarray}
In our work, we adopt $\theta \approx -10.7^\circ $. As shown in Table~\ref{Tab:couplings}, this yields a coupling ratio  $|g_{\kappa K \pi^{\pm}} |/| g'_{\kappa K \eta}| \approx 5.4$. 



 \bibliography{bibfile}

\end{document}